\documentclass[intlimits,twoside,a4paper]{article}

\usepackage{amsmath,amssymb}
\usepackage{graphicx}
\usepackage{wrapfig}

\usepackage[T2A]{fontenc}
\usepackage[cp1251]{inputenc}

\usepackage[eqsecnum]{cmpj2}
\issue{2015}{18}{3}{33706}
\doinumber{10.5488/CMP.18.33706}

\title[Self-consistent approach to magnetic ordering]%
{Self-consistent approach to magnetic ordering and excited site occupation processes \\ in a two-level system}
\author[M.A. Korynevskii, V.B. Solovyan]{M.A. Korynevskii\refaddr{label1,label2,label3},
        V.B. Solovyan\refaddr{label1}}
\addresses{
\addr{label1} Institute for Condensed Matter Physics of the National Academy of Sciences of Ukraine, \\ 1~Svientsitskii St., 79011 Lviv, Ukraine
\addr{label2} Lviv Polytechnic National University, 12~Bandera St., 79013 Lviv, Ukraine
\addr{label3} Institute of Physics of the University of Szczecin, 15~Wielkopolska St., 70451 Szczecin, Poland
}

\date{Received March 26, 2015, in final form July 1, 2015}
\authorcopyright{M.A. Korynevskii, V.B. Solovyan, 2015}

\newcommand{\no}{\nonumber}
\newcommand{\non}{\nonumber \\}

\newcommand{\lp}{\left (}
\newcommand{\rp}{\right )}
\newcommand{\lb}{\left [}
\newcommand{\rb}{\right ]}
\newcommand{\be}{\begin{equation}}
\newcommand{\ee}{\end{equation}}
\newcommand{\bea}{\begin{eqnarray}}
\newcommand{\eea}{\end{eqnarray}}

\begin{document}

\maketitle

\begin{abstract}
Ferromagnetic ordering in a two-level partially excited system is studied in detail.
Magnitudes of magnetization (magnetic order parameter)
and lattice ordering (excited level occupation number)
are calculated self-consistently. The influence of an external magnetic field and excited level gap on the ferromagnetic phase transition
is discussed.
\pacs 75.10.Hk, 75.30.Cr
\end{abstract}

\section{Introduction}

In numerous instances of real physical systems, the problem of ferromagnetic (or ferroelectric) ordering strongly depends on the density of magnetic (or dipole) particles in a medium. Such a situation is observed in different types of materials, namely in solids and liquids.
Particularly for solids, the irregularity in the position of magnetic particles is determined by a type of the sample preparation (slow or rapid cooling), and requires the use of equilibrium or nonequilibrium methods of statistics to  investigate them.

On the other hand, the irregularity of a system can be created by the excitation radiation or by heating in such a way that only a part of the total number of particles moves to the overlying energy levels. In case of constant external radiation (temperature), such a system is equilibrium and stable. The degree of occupation of the ground and excited states depends on their mutual energy distance and the intensity of radiation (thermostat temperature). The exchange interaction between the magnetic particles (in the ground or excited states) enables magnetic ordering processes in a  system. Their intensity essentially depends on the correlations between ground-ground, ground-excited and excited-excited states of particles.

Both an external magnetic field and radiation (thermostat temperature) form self-consistent thermodynamic  states of a system with a certain magnetization and an excited state occupation. For physically-critical applications, the intensity of external radiation (pump) or thermostat temperature may be considered as a constant.

A two-level model system can be used to investigate such a system and calculate its characteristics. Two-level systems are relatively simple, they are generally well studied (see \cite{ref1,ref2,ref3,ref4}) and form the basis for the microscopic mechanisms description of different physical phenomena and processes, such as non-linear optics, lasers, Josephson junctions, Kondo scattering, glasses, etc. \cite{ref5,ref6,ref7,ref8,ref9}. We shall regard the capability of spin particles of occupying one of the two levels: the ground and excited one with a certain probability dependent on the temperature and interaction between particles. A special attention is paid herein to the analysis of a correlation between spin orientation and occupation processes, as well as to the role of the external magnetic field and interlevel distance in the stable state of a system formation.

\section{Hamiltonian and free energy}

We consider the $N$-site lattice system of spin-like particles in each site which is capable of occupying one of the two quantum states, i.e., ground state and the excited one. The Hamiltonian of such a system in the external field $\gamma$, considering the pair exchange interactions between the particles, can be presented in the form:
\be
\label{eq2.1}
\hat H = - \gamma \sum_{i=1}^{N}\hat S_i -
\frac{1}{2} \sum_{i\neq j=1}^{N} \sum_{\lambda,\nu=1}^2
J_{ij}^{\lambda\nu} \hat C_i^{(\lambda)} \hat C_j^{(\nu)} \hat S_i  \hat S_j + \sum^N_{i=1} \sum^2_{\lambda=1} \epsilon_\lambda \hat C_i^{(\lambda)}.
\ee
Here, $\hat S_i = \hat S_i^z$ is $z$ Pauli matrix; $\hat C_i^{(\lambda)}$ is the operator of the number of particles in the $i$-th site in the quantum state $\lambda$; $J_{ij}^{\lambda\nu}$ is an integral of the exchange interaction between spin particles in the $i$-th and $j$-th sites of the crystalline lattice;
$\epsilon_\lambda$ is a configuration part of the energy per one particle in the $\lambda$ state.
Since the total number of particles coincides with $N$,
\be\label{eq2.2}
\sum^N_{i=1} \sum^2_{\lambda=1} \hat C_i^{(\lambda)} = N \hat I,
\ee
\[
\hat C_i^{(1)} = \left( \begin{array}{cc} 1 & 0\\0 & 0 \end{array}\right), \qquad
\hat C_i^{(2)} = \left( \begin{array}{cc} 0 & 0\\0 & 1 \end{array}\right), \qquad
\hat I = \left( \begin{array}{cc} 1 & 0\\0 & 1 \end{array}\right),
\qquad \hat S_i = \left( \begin{array}{cc} 1 & 0\\0 & -1 \end{array}\right),
\]
we can pass from two operators $\hat C_i^{(1)}$ and $\hat C_i^{(2)}$ to the unique operator $\hat C_i = \hat C_i^{(2)}$, which coincides with the excited particle number operator. The eigenvalues of  operator $\hat C_i$  are:
\be
\label{eq2.3}
C_i = \left\{ \begin{array}{ll}
0, \qquad \text{when a particle in the $i$-th site is in the ground state},\\
1, \qquad \text{when a particle in the $i$-th site is in the excited state}.
\end{array}\right.
\ee

The orthonormal set of the wave functions for each site of the lattice is of the form of the product of configuration and spin components in the following way:
\be
\label{eq2.4}
  \left\{\psi_i^{(\lambda,s)}\right\} \equiv \left\{\varphi_i^{(\lambda)}\right\} \cdot  \left\{\chi_i^{(s)}\right\},
\ee
\be
\varphi_i^{(1)}= \left(\begin{array}{cc}1\\0\end{array}\right)_{c}, \quad \varphi_i^{(2)}= \left(\begin{array}{cc}0\\1\end{array}\right)_{c}; \qquad \qquad
\chi_i^{(1)}= \left(\begin{array}{cc}1\\0\end{array}\right)_{s}, \quad \chi_i^{(-1)}= \left(\begin{array}{cc}0\\1\end{array}\right)_{s}. \nonumber
\ee
Thus, in reality we have a four-state situation.

It is well-known \cite{ref10,ref11,ref12} that for atoms, the wave functions of excited electron states are much broader than the other ones. As a result, the exchange integrals $J_{ij}^{11}$  between a pair of the nearest electrons being in the ground state (similarly between a particle in the ground state and the second one in the excited state $J_{ij}^{12}$) can be regarded as small compared with the exchange integral $J_{ij}^{22}$ between both particles in excited states.

As a zero approximation for exchange interaction in expression (\ref{eq2.1}), we take $J_{ij}^{11}$ and $J_{ij}^{12}$ negligibly small compared with $J_{ij}^{22}$. This is not a principal limitation, but makes it possible to use only one interaction parameter $J_{ij}^{22}$. Such an approach is very close to Vonsovskii $s$--$d$ exchange model in the theory of magnetism \cite{ref13,ref14}, according to which $s$-electrons are responsible for electric conductivity while $d$-electrons form magnetic properties of solids. The comparison of exchange integrals for ground and excited states for rare-earth metals can be also found in \cite{ref15}.

Following the above mentioned restrictions and taking the configurational part of energy of particles
 in the ground state equal to zero, the model Hamiltonian (\ref{eq2.1}) takes the form:
\be
\label{eq2.5}
\hat H = - \gamma \sum_{i=1}^{N}\hat S_i - \frac{1}{2} \sum_{i\neq j=1}^{N} J_{ij}^{22} \hat C_i  \hat C_j \hat S_i \hat S_j + \epsilon_0
\sum_{i=1}^{N} \hat C_i\,.
\ee
Here, $\epsilon_0\equiv\epsilon_2-\epsilon_1$ plays the role of energy gap between excited and ground states for a particle in each site, $J_{ij}^{22}>0$, which corresponds to ferromagnetic exchange coupling.

By certain features, the model described by the Hamiltonian (\ref{eq2.5}) is close to a well-known Blume-Emery-Griffits (BEG) model \cite{ref16} introduced to simulate the thermodynamic behaviour of $^3$He-$^4$He mixtures and to numerous modifications of BEG model used for interpretation of diluted magnets properties \cite{ref17,ref18,ref19,ref20,ref21,ref22,ref23}. The papers in this field of investigation are generally based on the use of the Ising Hamiltonian with a spin equal to unity as well as on some assumption concerning the character of randomness for magnetic particles in a lattice. Briefly speaking, those models are three-state site models due to the  three possible projections of $S^z$-operator. Quite different physical situation is observed in our investigation, namely a physical interpretation of expression (\ref{eq2.5}) comes to two-level, four-state site model (two spin-orientations for the particle in the ground state and two spin-orientations for this particle in the excited state). Temperature helps to increase the excited state occupation but destroys spontaneous magnetization caused by  the  external field and the exchange interaction between spin particles in the excited states. Such an approach, in our opinion, is new and quite interesting. Moreover, both quantities $c$ and $s$ being physically different are  calculated here  self-consistently.

The method of self-consistent field approximation is used to calculate statistical and thermodynamical properties of the system described by the Hamiltonian (\ref{eq2.5}). From the physical point of view, this approximation corresponds to the neglect of square fluctuations of the identical physical values for a pair of different sites in a  crystalline lattice, the same being true for different physical values in the common site. It is well-known that this approximation is valid in a wide range of temperatures and external fields except for the immediate phase transition point neighbourhood \cite{ref24,ref25}. Thus, we accept the following expressions for the products of the operators:
\begin{align}
&
\hat{C_i} \hat{C_j} = \langle\hat{C_i}\rangle \hat{C_j} + \hat
 C_i \langle\hat{C_j}\rangle
- \langle\hat{C_i}\rangle \langle\hat{C_j}\rangle,\non
&
\label{eq2.6}
\hat S_i \hat S_j = \langle \hat S_i\rangle \hat S_j + \hat S_i \langle \hat S_j\rangle - \langle \hat S_i\rangle \langle \hat S_j\rangle, \\
&
\hat{C_i} \hat{S_i} = \langle\hat{C_i}\rangle \hat{S_i} + \hat C_i \langle\hat{S_i}\rangle
- \langle\hat{C_i}\rangle \langle\hat{S_i}\rangle,\no
\end{align}
where $ \langle\hat{C_i}\rangle \equiv c $ and $\langle\hat{S_i}\rangle \equiv s$ are mean values of $\hat C_i$ and $\hat S_i$ operators, correspondingly, according to the Gibbs distribution based on the Hamiltonian (\ref{eq2.5}) and expressions (\ref{eq2.6}).

Being limited in the summation over the sites $i$, $j$ only to the nearest neighbours
($ \sum_j J_{ij}^{22}=X_vJ$, $X_v$ is the number of the nearest neighbours, $J$ is an interaction constant), we obtain for $\hat H$ the following expression:
\[
\hat H = \sum_{i=1}^N \hat H_i = N \hat H_i\,,
\]
\be
\label{eq2.7}
\hat H_i = - \gamma \hat S_i - c s X_v J \left( c \hat S_i + s \hat C_i \right) + \epsilon_0
\hat C_i  + \frac{3}{2} c^2 s^2 X_v J.
\ee
Thus, the Hamiltonian (\ref{eq2.7}) describes the $N$-particle system with one spin particle in each site. At the same time, each particle staying at one of the two energy levels can take up one of the four states. The total number of particles in a system is constant, so canonical Gibbs ensemble may be used.

Contrary to the well-known situations for the instance of independent subsystems: 1) pure or diluted magnetic system ($c= \textrm{const}$), or 2) simple lattice-gas system ($s=\text{const}$), the Hamiltonian (\ref{eq2.7}) describes both binding subsystems, and the values of $c$ and $s$ are in a close connection depending on temperature $T$, external field $\gamma$ and energy gap $\epsilon_0$.

A partition function of the system described by the Hamiltonian (\ref{eq2.7}) is presented as:
\be
\label{eq2.8}
Z = \textrm{Tr} \left[{\exp(-\beta \hat H}) \right] = \left\{ \textrm{Tr} \left[ \exp({-\beta\hat H_i})\right]\right\}^N.
\ee
Here, $\beta=(k_\textrm{B}T)^{-1}$, $k_\textrm{B}$ is the Boltzmann constant, $T$ is the absolute temperature, and thus the  $\textrm{Tr}$ operation should be performed based on the wave functions (\ref{eq2.4}).

After performing this operation we obtain the following expression for the free energy $F$ per one site:
\be
\label{eq2.9}
F = - \frac{1}{\beta} \ln \left\{   2 \cosh \left[ \beta\left(c^2 s X_v J + \gamma \right)\right]   \right\}
- \frac{1}{\beta}\ln \left\{ 1 + \exp \left[ \beta(c s^2 X_vJ - \epsilon_0) \right] \right\} + \frac{3}{2} c^2 s^2 X_v J.
\ee

\section{Order parameters $c$ and $s$}

The free energy in the form of (\ref{eq2.9}) is the function of three independent parameters:
temperature $T$, external field $\gamma$ and excited energy gap $\epsilon_0$. Using a self-consistent field approximation, we introduced two order parameters
$c$ and $s$, which as the functions of $T$, $\gamma$, $\epsilon_0$ must satisfy the conditions of thermodynamic stability in the form \cite{ref26,ref27}:
\be
\label{eq3.1}
\frac{\partial F}{\partial a_i} = 0, \qquad i = 1, 2, 3, \ldots \ ,
\ee
where $a_1=c$ and $a_2=s$ in our instance. On the other hand, a system entropy $S$ and magnetization per one site $s$ can be calculated as partial derivatives:
\be
\label{eq3.2}
S = - \frac{\partial F}{\partial T}\,,
\ee
\be
\label{eq3.3}
s =  - \frac{\partial F}{\partial \gamma}\,.
\ee
Undoubtedly, equation (\ref{eq3.1}) for $i=2$ and equation (\ref{eq3.3}) coincide.

Based on the equations (\ref{eq3.1}), we get:
\bea
&&
3 c s - 2 c \tanh \left[ \beta \left( c^2 s X_v J + \gamma \right)\right] -
s \left\{ 1 + \exp \left[-\beta(cs^2X_vJ-\epsilon_0)\right] \right\}^{-1}=0,\non
&&
\label{eq3.4}
3 c s -  c \tanh \left[ \beta \left( c^2 s X_v J + \gamma \right)\right] -
2s \left\{ 1 + \exp \left[-\beta(cs^2X_vJ-\epsilon_0)\right]\right\}^{-1}=0.
\eea
Combining these two equations we obtain for $c$ and $s$ the system of equations in a more convenient form:
\bea
&&
c = \frac{1}{2} \left\{ 1 + \tanh \lb \frac{\beta}{2}  \lp c s^2 X_v J -\epsilon_0 \rp\rb \right\},\non
&&\label{eq3.5}
s = \tanh \left[ \beta \lp c^2 s X_v J + \gamma \rp\right].
\eea

It can be shown that for arbitrary values of $X_v J$, the following relationship between $c$ and $s$ takes place:
\be
\label{eq3.6}
\exp \left[ 2\beta(c\epsilon_0+s \gamma)\right] = \left( \frac{1+s}{1-s}\right)^s \left( \frac{1-c}{c}\right)^{2c},
\ee
which at $\gamma=0$ and $\epsilon_0=0$ transforms into a simpler form:
\be
\label{eq3.7}
\left( \frac{1-s}{1+s}\right)^s = \left( \frac{1-c}{c}\right)^{2c}.
\ee
Equations (\ref{eq3.6}),~(\ref{eq3.7}) show an unambiguous relation between $c$ and $s$ for different values of external parameters $T$, $\gamma$, $\epsilon_0$.

\begin{figure}[!t]
\centerline{
\includegraphics[width=0.44\textwidth]{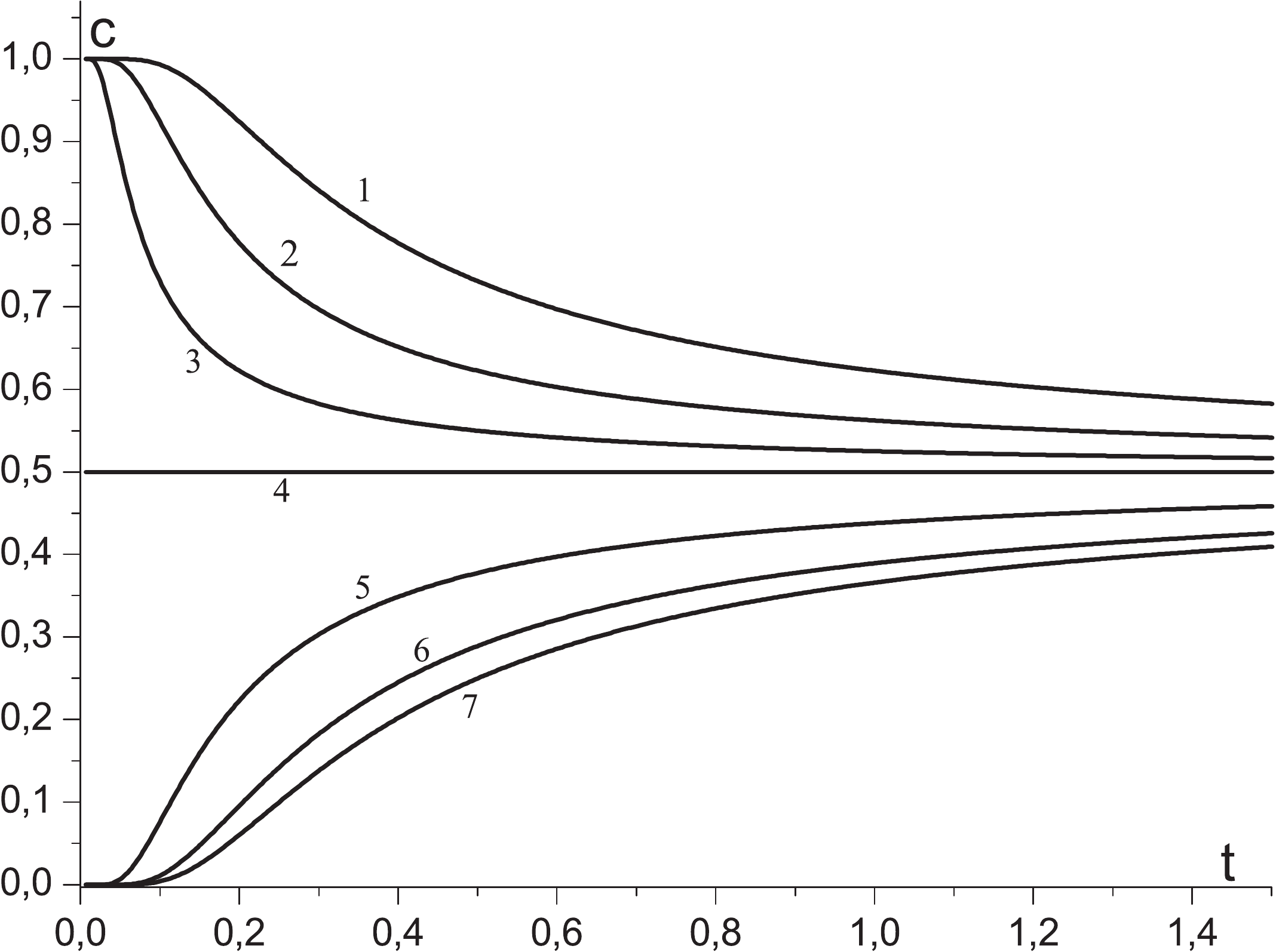}
\qquad
\includegraphics[width=0.44\textwidth]{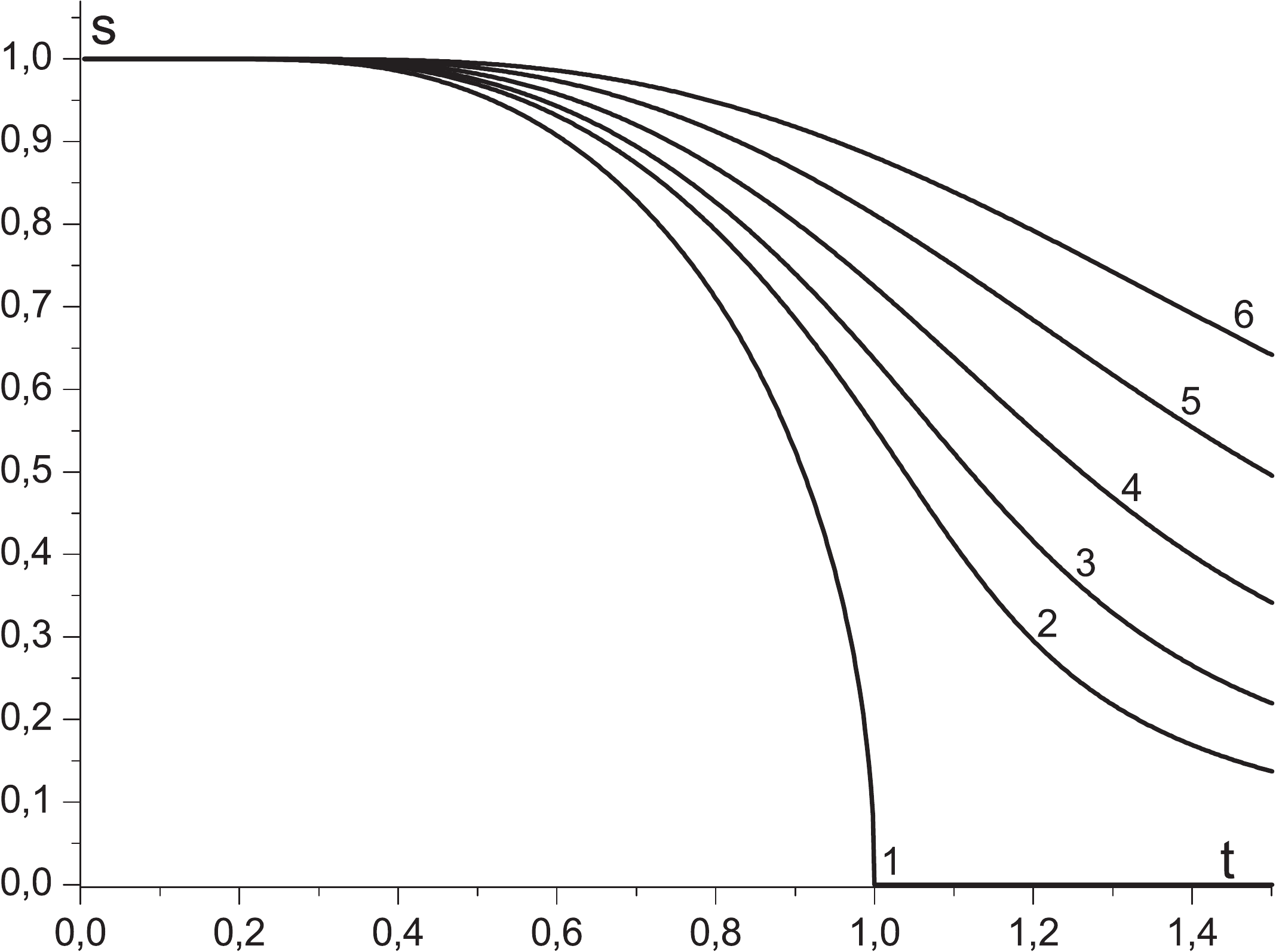}
}
\centerline{(a) \hspace{0.46\textwidth} (b)}
\caption{Temperature dependencies of independent occupation number
[figure~\ref{fig1}~(a), curves:  1~--- $\epsilon=-0.5$, 2~--- $\epsilon=-0.25$, 3~--- $\epsilon=-0.1$, 4~--- $\epsilon=0$, 5~--- $\epsilon=0.25$, 6~--- $\epsilon=0.45$, 7~--- $\epsilon=0.55$]  and magnetic order parameter [figure~\ref{fig1}~(b), curves: 1~--- $h=0$, 2~--- $h=0.07$, 3~--- $h=0.115\dots$,  4~--- $h=0.192\dots$, 5~--- $h=0.318\dots$, 6~--- $h=0.5$, ($c_0=1$)].}
\label{fig1}
\end{figure}

We start the analysis of equations (\ref{eq3.5})  from the situation where  the occupation process and magnetic ordering are completely independent. In other words, the occupation of the excited level $c$ is determined by temperature only (the effect of magnetic ordering is missing, $s=0$). In this instance, according to (\ref{eq3.5}), we obtain:
\be
\label{eq3.8}
c = \left[ 1 + \exp \left(\beta\epsilon_0 \right)\right]^{-1},
\ee
and, correspondingly,
\be
\label{eq3.9}
s = \tanh \left[ \beta \left( c_0^2 s X_v J + \gamma \right)\right],
\ee
where $c_0$ is independent of temperature and has got a fixed (for example by external radiation) concentration of particles in the excited states. The reduced temperature $t = T (X_v J)^{-1}$ dependencies of $c$ [figure~\ref{fig1}~(a)] and $s$ [figure~\ref{fig1}~(b)] for  different values of the reduced energy gap  $\epsilon = \epsilon_0(X_v J)^{-1}$ and the reduced external field $h = \gamma(X_v J)^{-1}$ are presented in figure~\ref{fig1}.  One can observe a smooth behaviour of both $c$ and $s$ for any  $h$, $\epsilon$. The curves for magnetization are characteristic of the instance of a second order phase transition.

\begin{figure}[!t]
\centerline{
\includegraphics[width=0.41\textwidth]{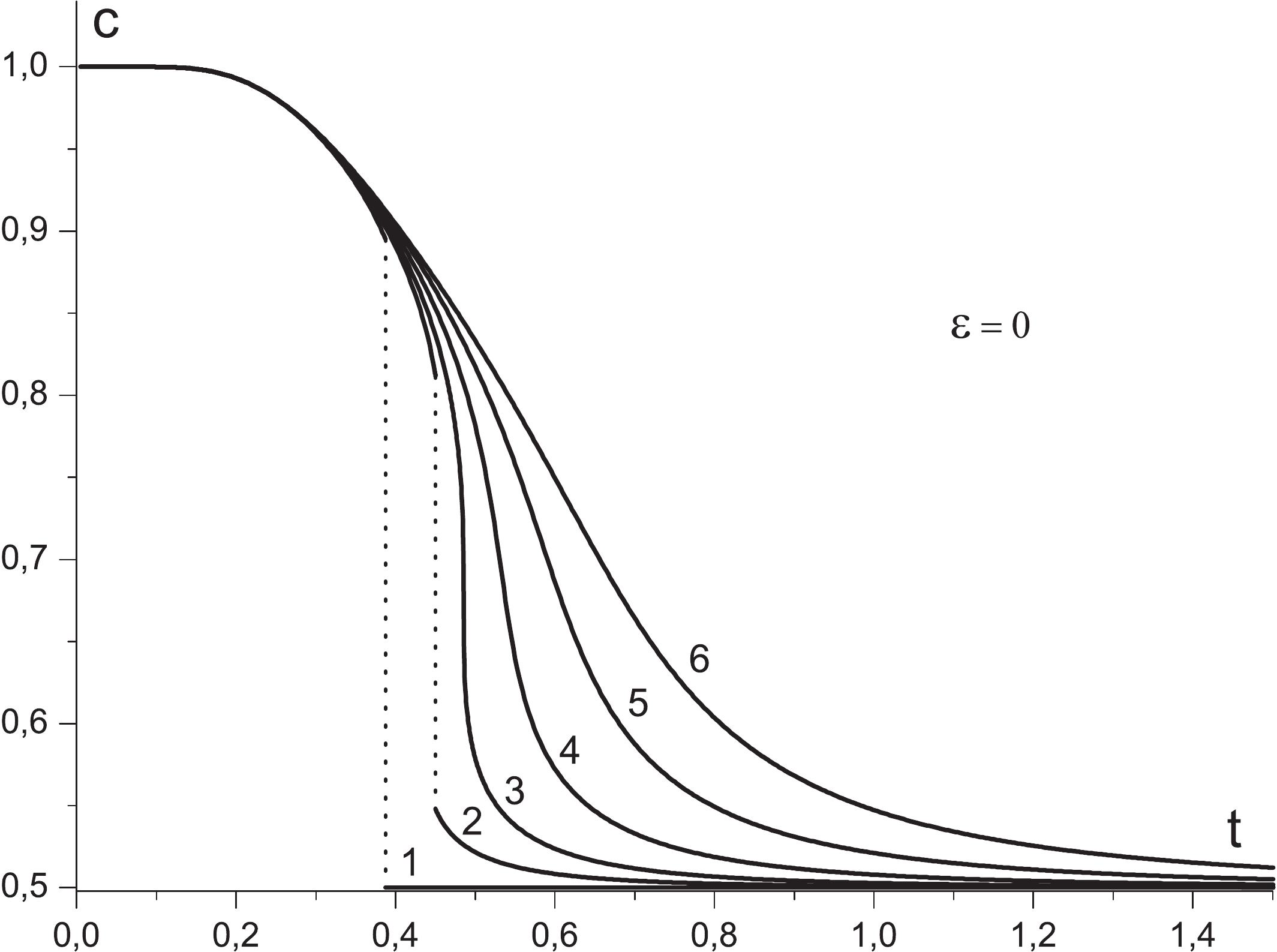}
\qquad
\includegraphics[width=0.41\textwidth]{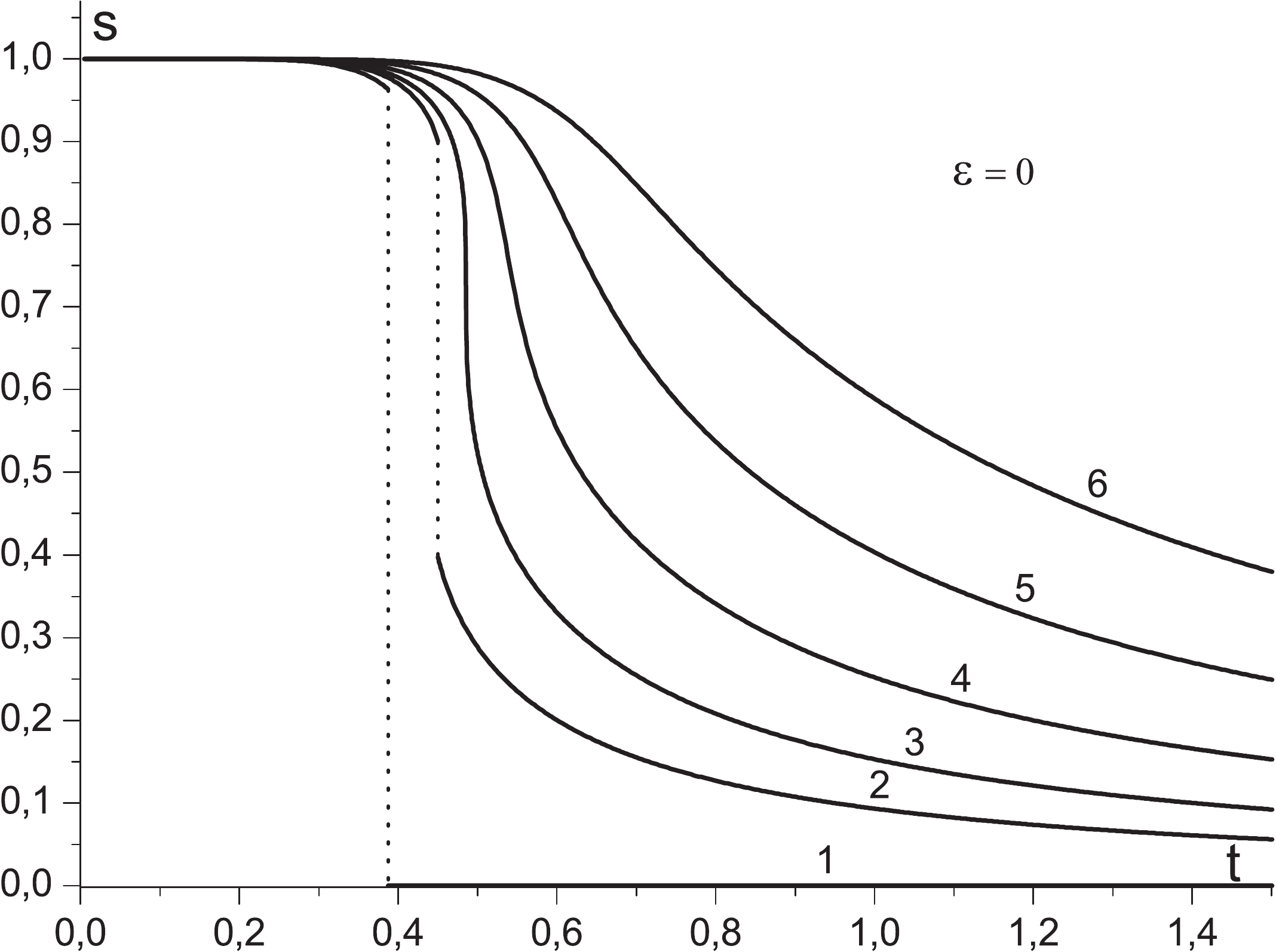}
}\centerline{(a) \hspace{0.45\textwidth} (e)}
\centerline{
\includegraphics[width=0.41\textwidth]{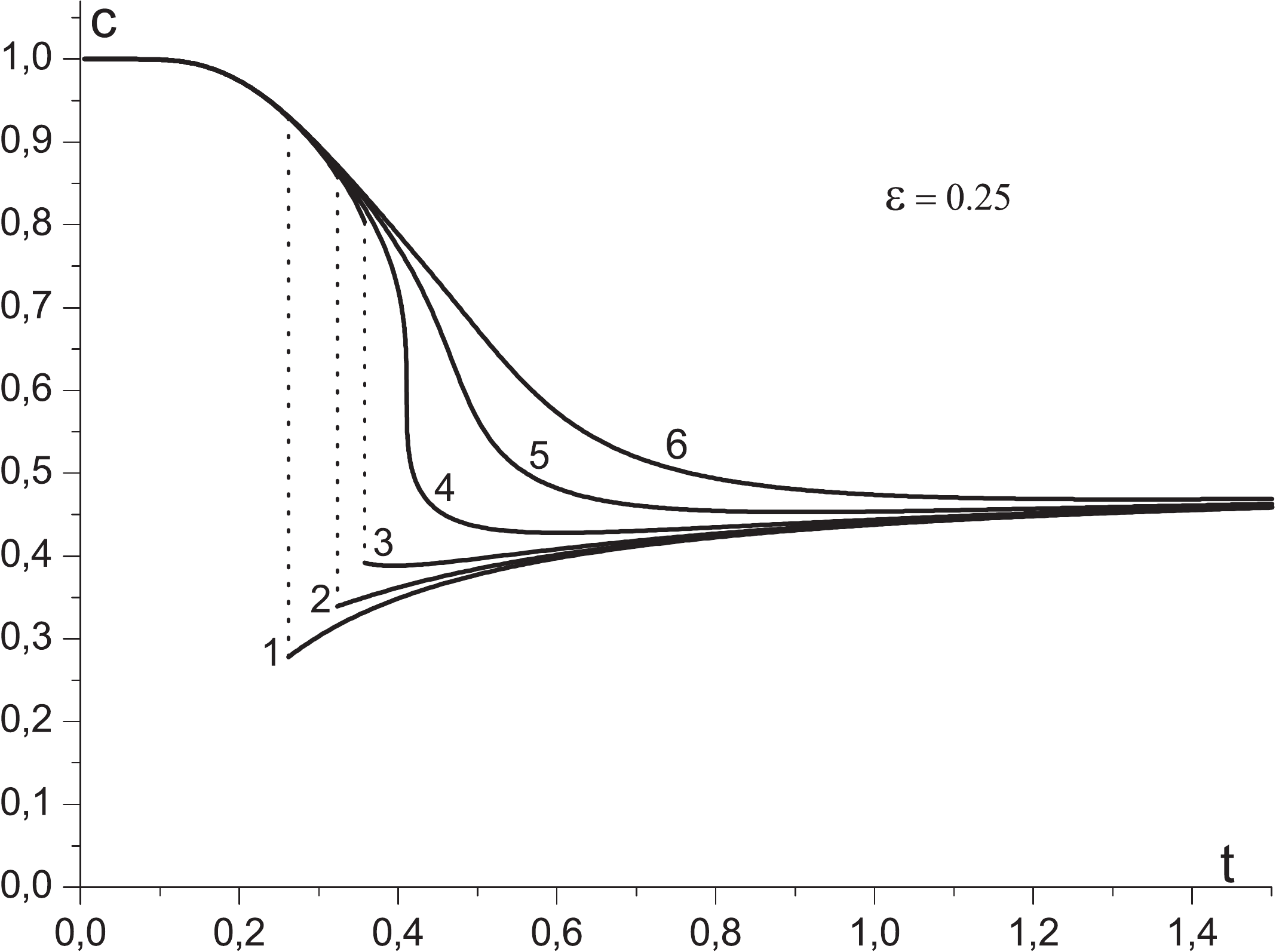}
\qquad
\includegraphics[width=0.41\textwidth]{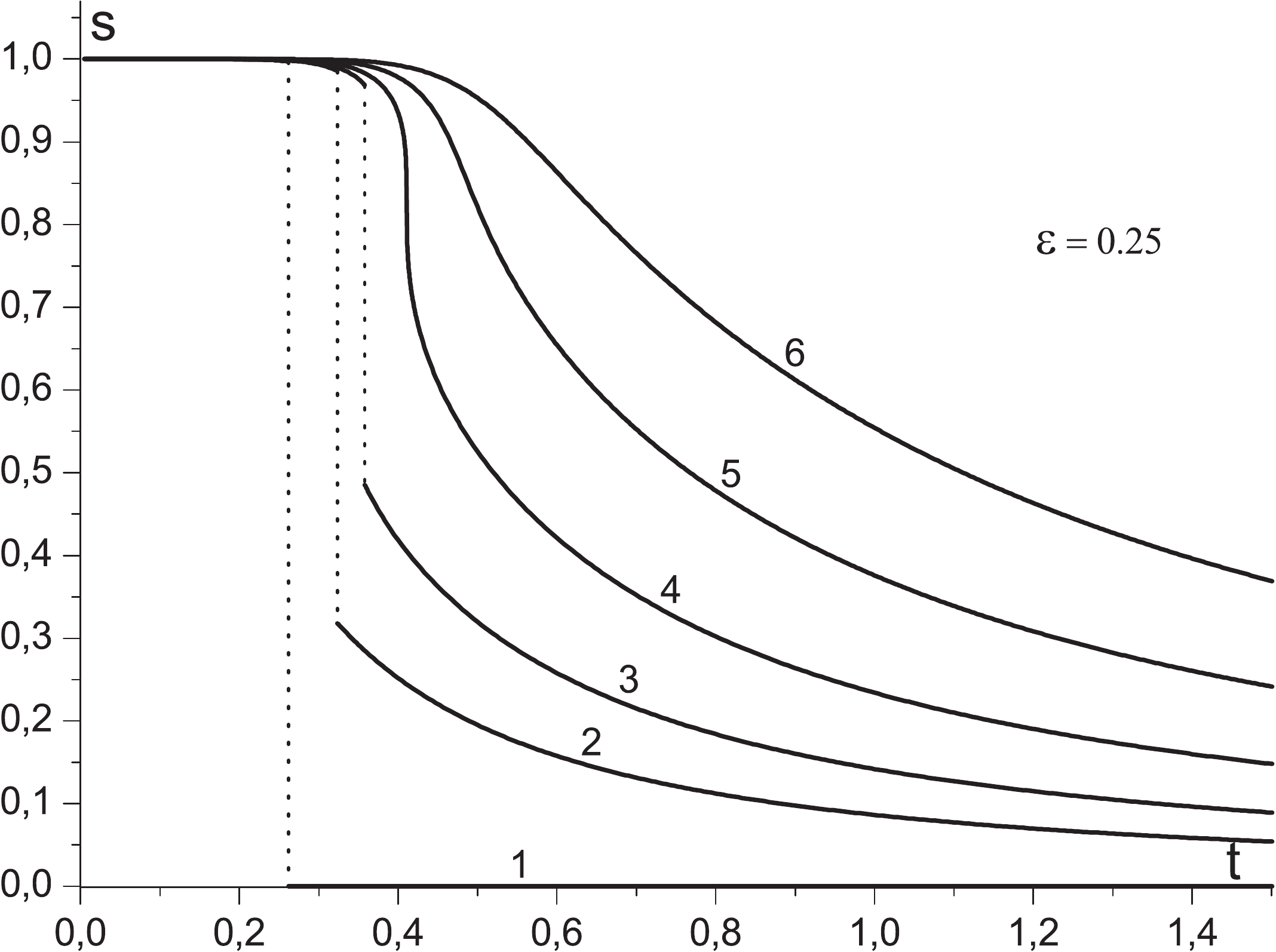}
}\centerline{(b) \hspace{0.45\textwidth} (f)}
\centerline{
\includegraphics[width=0.41\textwidth]{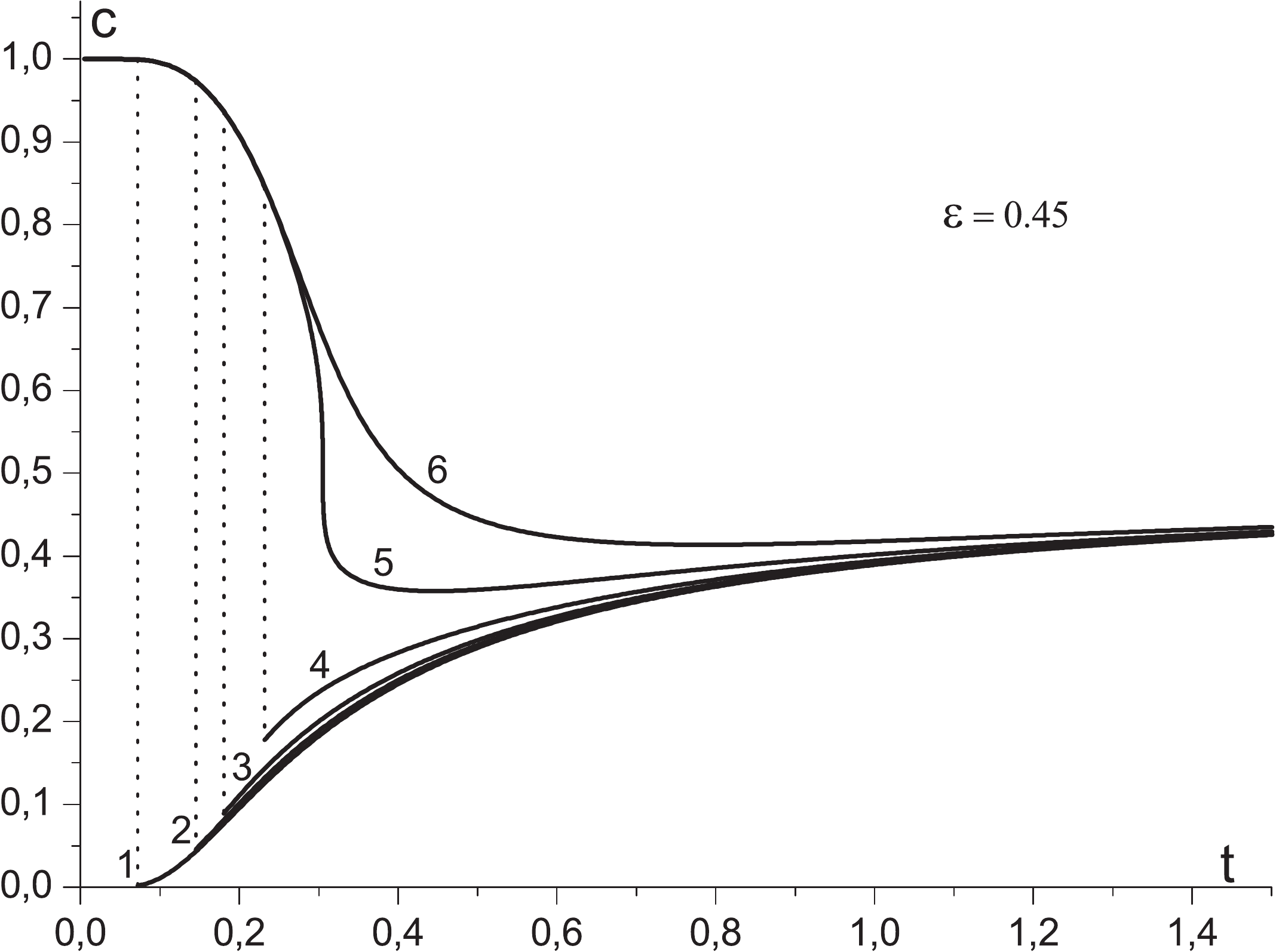}
\qquad
\includegraphics[width=0.41\textwidth]{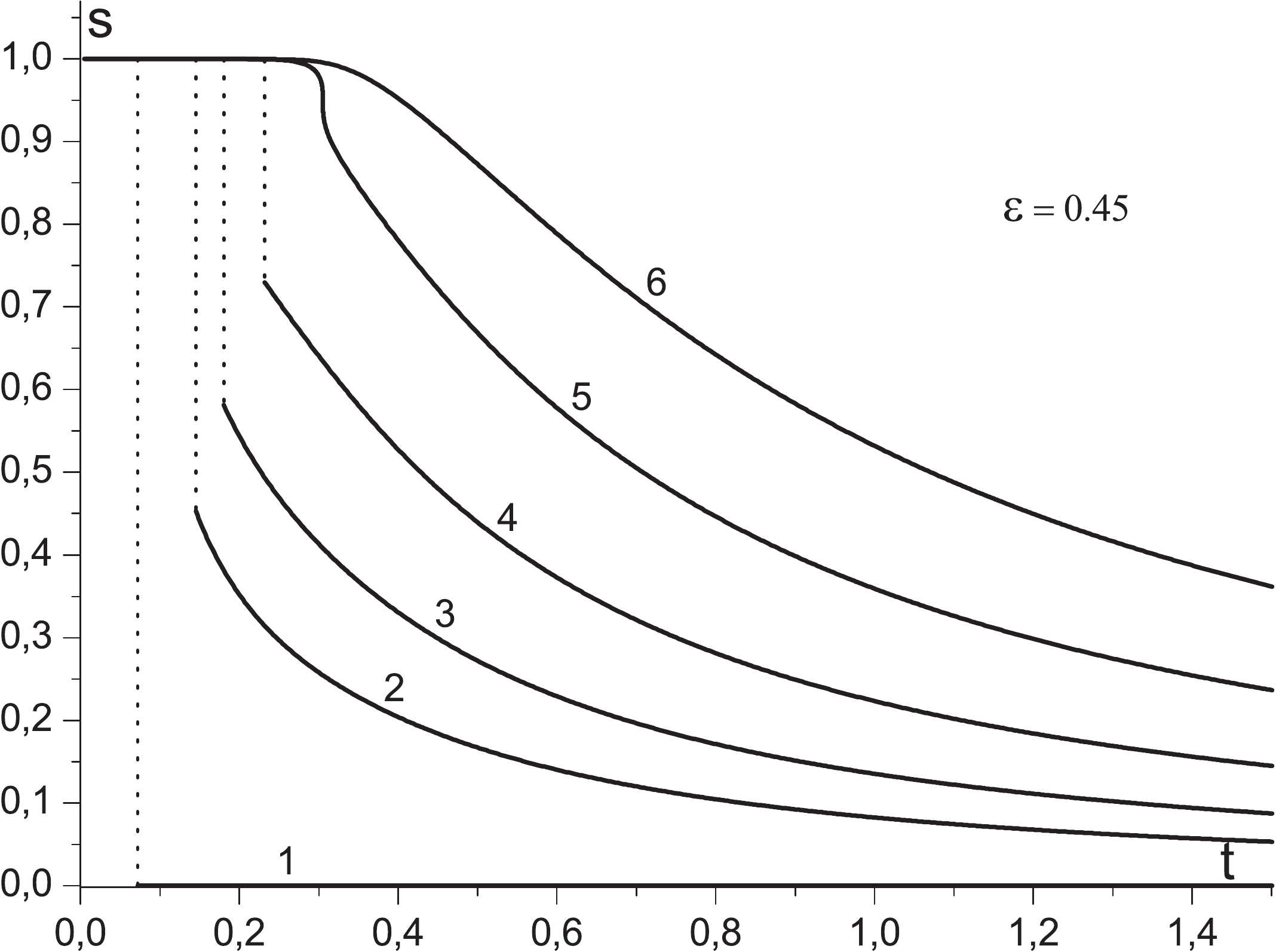}
}\centerline{(c) \hspace{0.45\textwidth} (g)}
\centerline{
\includegraphics[width=0.41\textwidth]{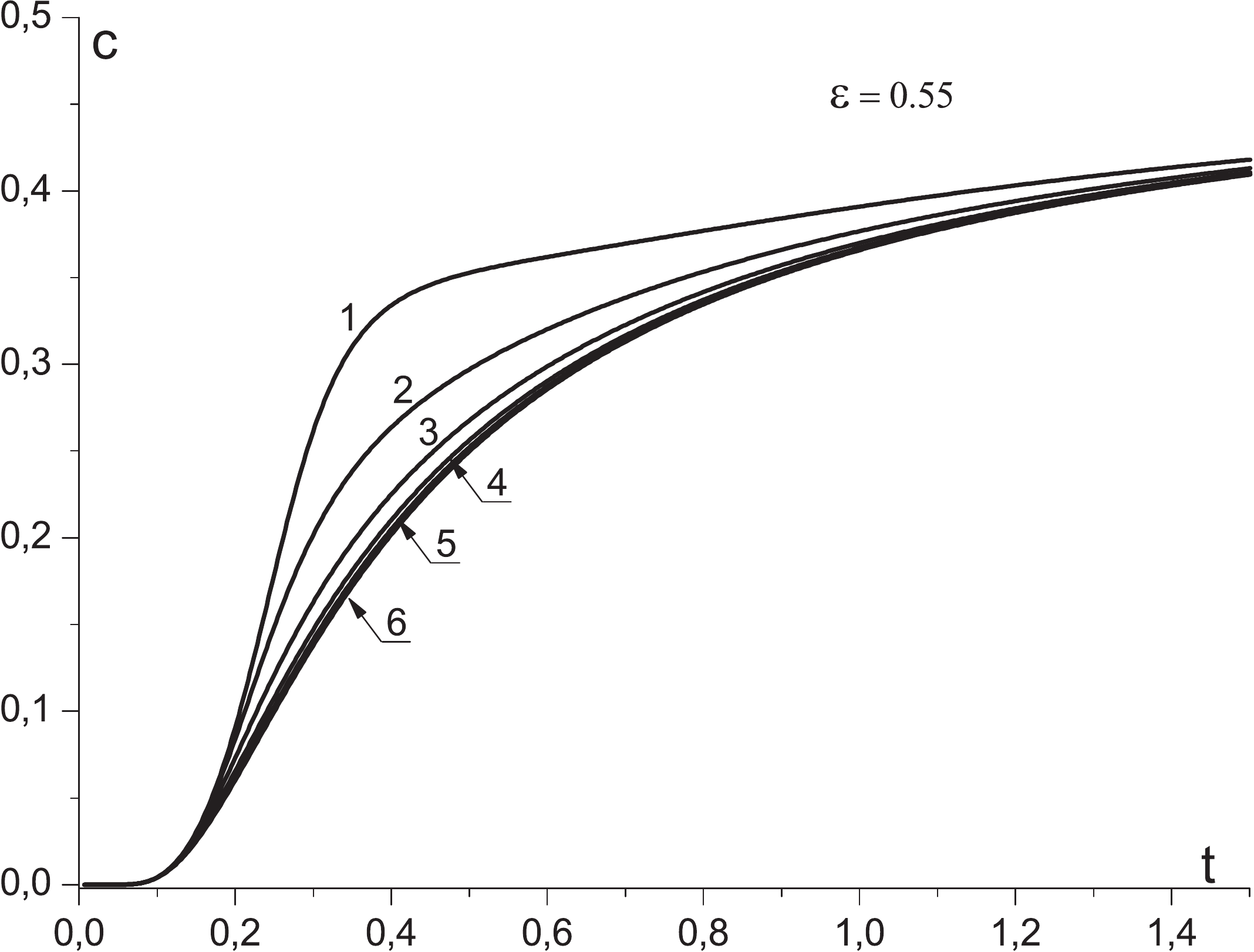}
\qquad
\includegraphics[width=0.41\textwidth]{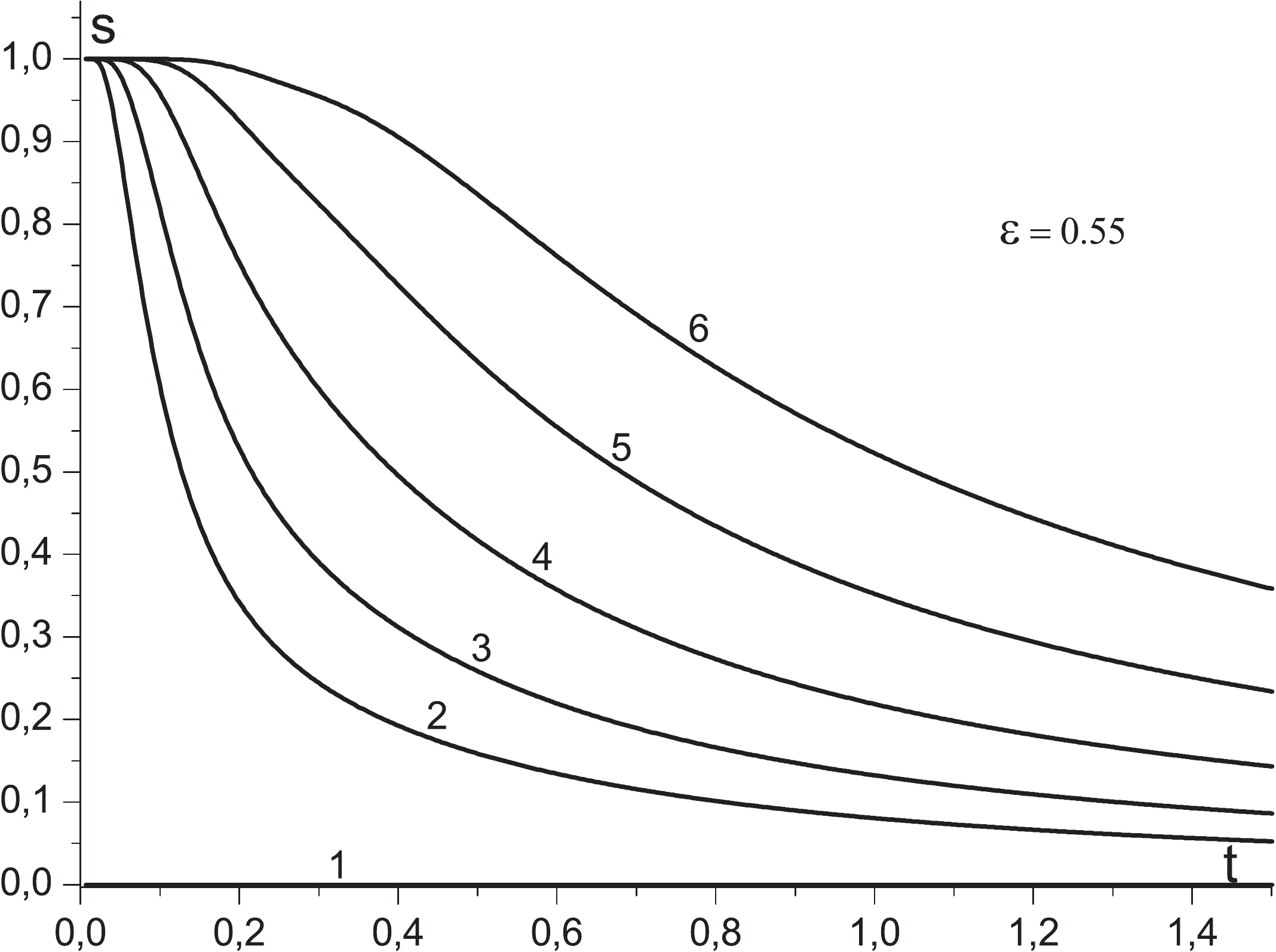}
}\centerline{(d) \hspace{0.45\textwidth} (h)}
\caption{Temperature dependencies for occupation number  [figures~\ref{fig2}~(a)--(d)] and magnetic order parameter [figures~\ref{fig2}~(e)--(h)] at different energy gaps $\epsilon$ and external fields $h$ (curves: 1~--- $h=0$, 2~--- $h=0.07$, 3~--- $h=0.115\dots$,
4~--- $h=0.192\dots$, 5~--- $h=0.318\dots$, 6~--- $h = 0.50$).}
\label{fig2}
\end{figure}

Quite different is the situation when considering bounded system equations (\ref{eq3.5}) (see figure~\ref{fig2}). Now the correlation between two order parameters ($c$ and $s$) becomes significant and the role of ``external'' independent factors ($h, \epsilon$) is non-trivial.

For each value of energy gap $\epsilon < 0.5$, the curves for $c$ and $s$ demonstrate a smooth temperature behaviour for $h>h_\textrm{c}$ ($h_\textrm{c}$ dependent on $\epsilon$). At $h<h_\textrm{c}$,  $c$ and $s$ show jumps at a certain temperature. The values of those jumps decrease with an increase of  $h$. At $h=h_\textrm{c}$, this jump turns into zero [curves~3 in figures~\ref{fig2}~(a),(e), curves~4 in figures~\ref{fig2}~(b),(f), curves~5 in figures~\ref{fig2}~(c),(g)]. Consequently, for $h<h_\textrm{c}$, the first order magnetic phase transition takes place in the investigated system, at $h=h_\textrm{c}$ the order of phase transition changes into a second order (tricritical point) and for $h>h_\textrm{c}$ no magnetic phase transition is possible. The point $h_\textrm{c}$ divides the external fields into ``weak'' and ``strong'' ones \cite{ref27}. Those results are presented in figures~\ref{fig2}~(a)--(c) and figures~\ref{fig2}~(e)--(g).

For $\epsilon>0.5$ instance, only smooth curves $c$ and $s$ are observed. The role of the external field $h$ is quantitative, no critical $h_\textrm{c}$ was found [figures~\ref{fig2}~(d),(h)] and spontaneous magnetization is impossible.

A special role of $\epsilon$  parameter is connected with a strong correlation between the processes of the excited level occupation and spontaneous ordering of magnetic particles at this level. Heating facilitates the filling of the excited level while the ordered magnetic state decreases. The measure of those processes is the value of the system energy, i.e., the ratio between its growth in the excitation and a decrease due to a magnetic ordering. We observe a situation close to the phenomenon of percolation in the systems having a constant concentration of magnetic particles \cite{ref28}. However, the phenomenon of percolation in the system studied occurs self-consistently where the excited level occupation by magnetic particles depends on the degree of magnetization. That is why for the first order magnetic phase transition, a sharp decrease (dip) of the excited level occupation  is observed above the phase transition temperature point $T_\textrm{c}$ [see figures~\ref{fig2}~(a)--(d)].

Taking into account that magnetization of the system is represented by the following formula:
\be
\label{eq3.10}
\vec M = \vec\mu N s,
\ee
where $\vec\mu$ is a magnetic moment of an individual particle, based on the equations (\ref{eq3.5}), we have obtained the expression for generalized magnetic susceptibility (as a function of temperature and external magnetic field):
\bea
\label{eq3.11}
\chi &=& |\vec\mu| N \frac{Y_1}{X_v J(Y_1Y_2-4c^2s^2)}\,,
\\
\label{eq3.12}
Y_1 &=& \frac{1}{ \beta X_v J (1-c)c} - s^2,
\\
Y_2 &=& \frac{1}{ \beta X_v J(1-s^2)} - c^2 . \nonumber
\eea

\begin{figure}[!t]
\centerline{
\includegraphics[width=0.43\textwidth]{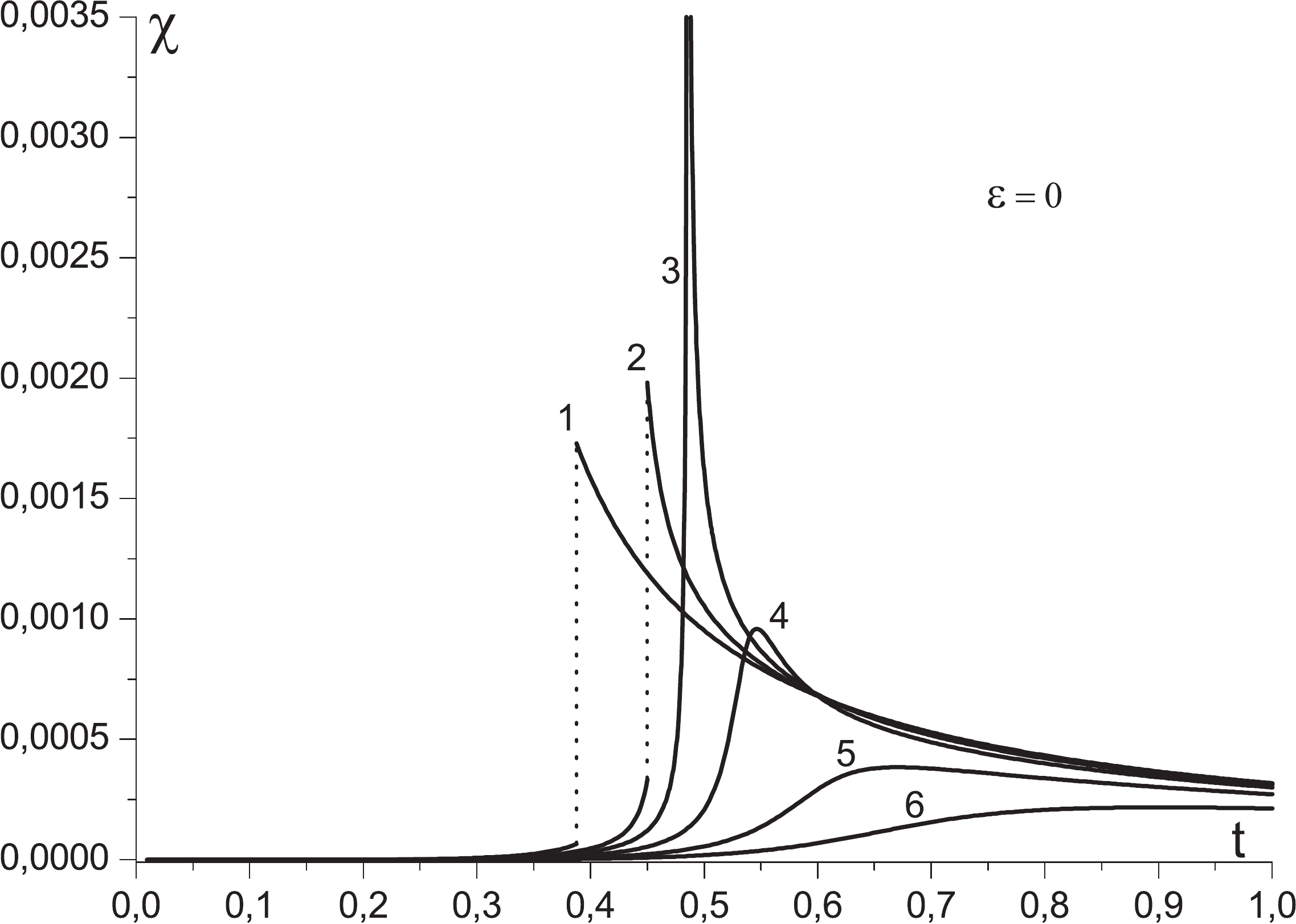}
\qquad
\includegraphics[width=0.43\textwidth]{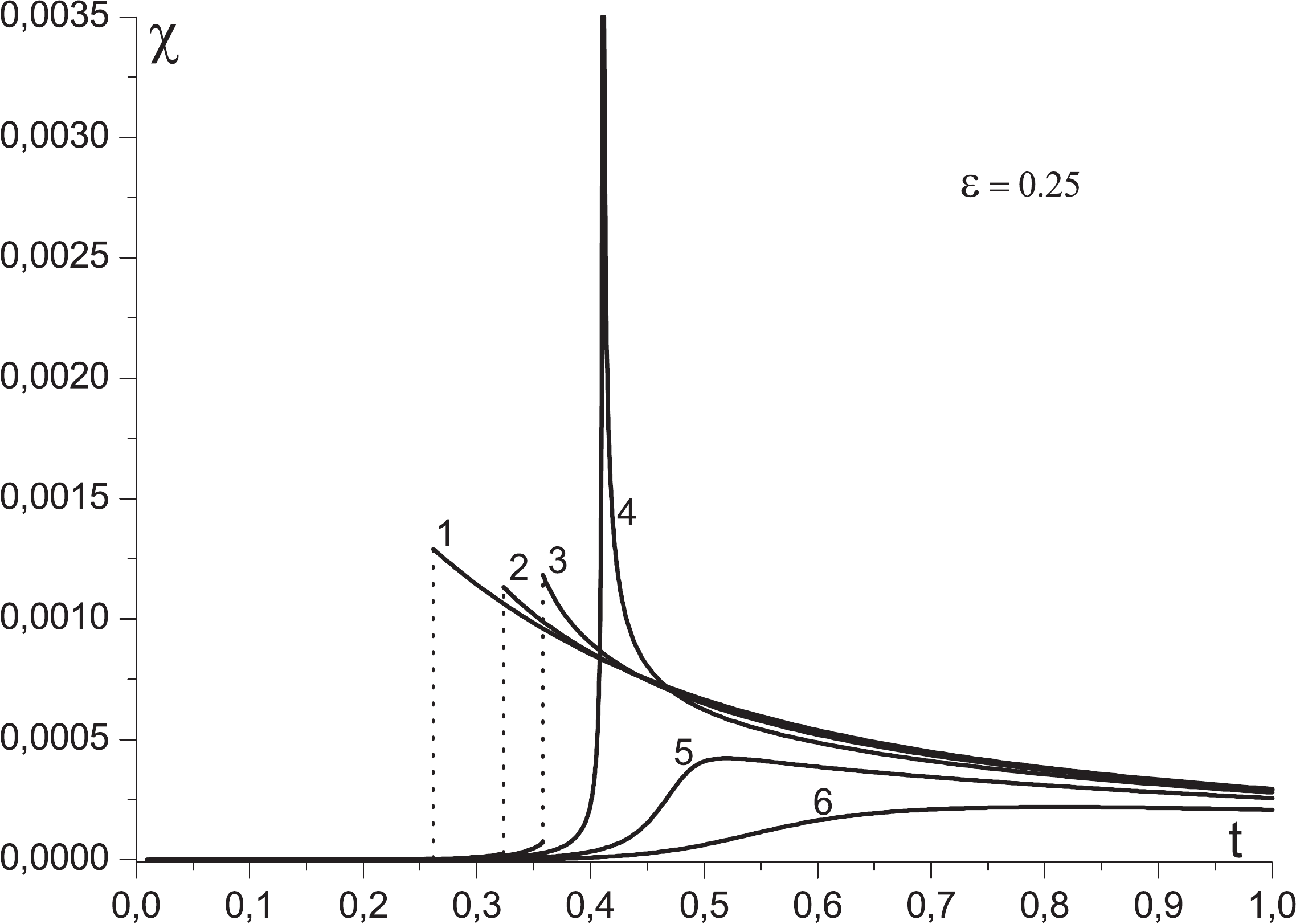}
}\centerline{(a) \hspace{0.45\textwidth} (b)}
\centerline{
\includegraphics[width=0.43\textwidth]{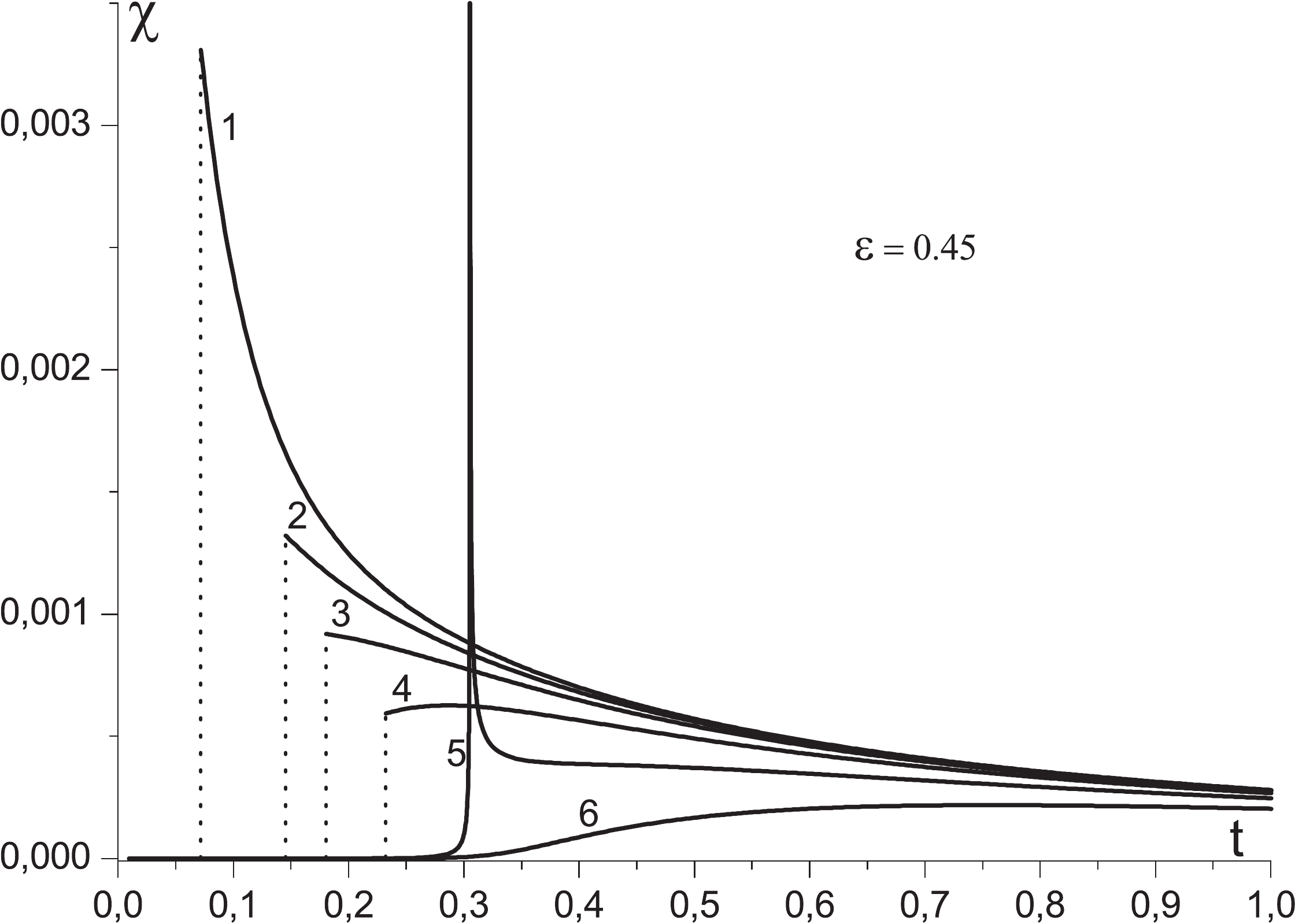}
\qquad
\includegraphics[width=0.43\textwidth]{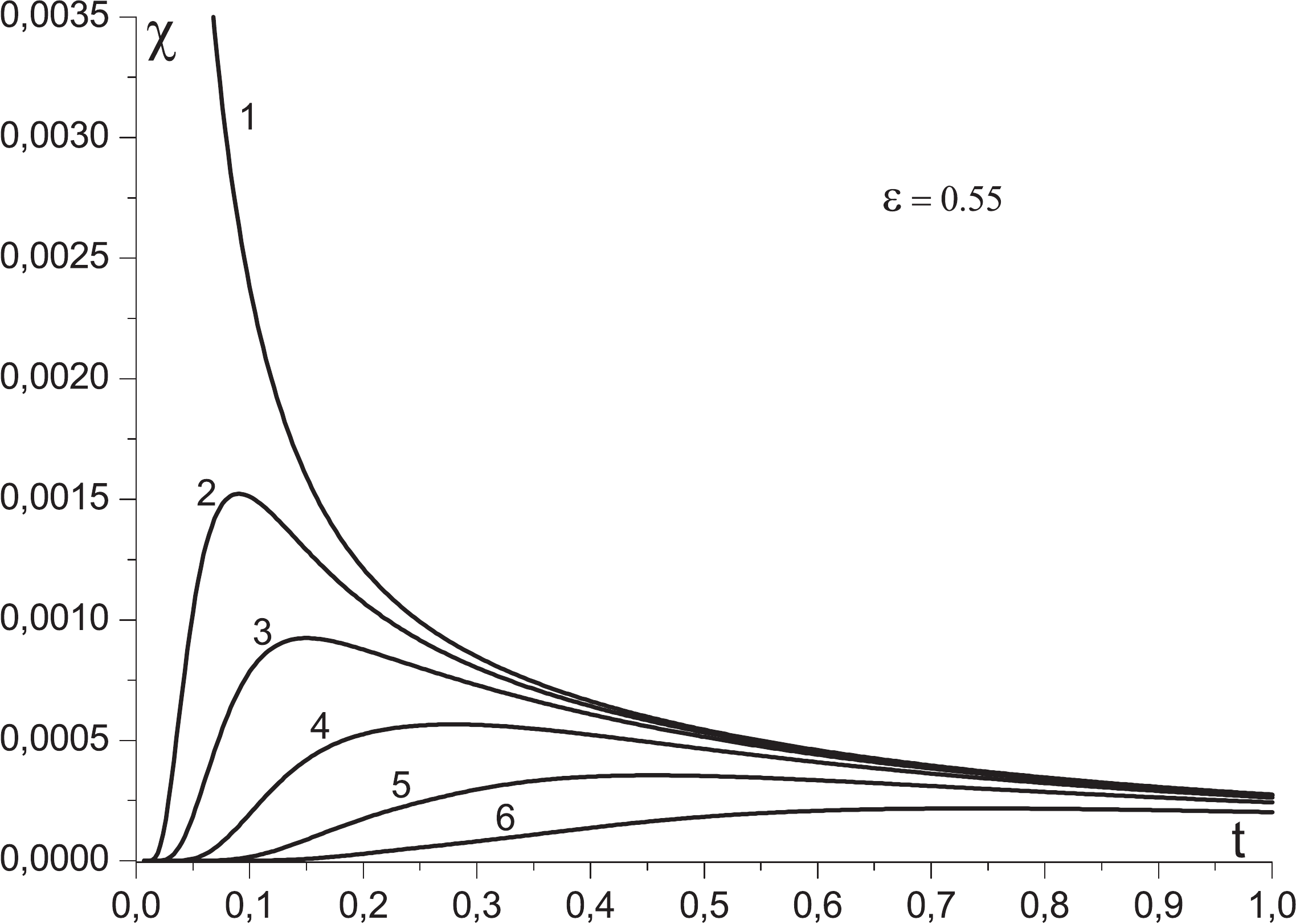}
}\centerline{(c) \hspace{0.45\textwidth} (d)}
\caption{
Temperature dependencies of generalized  magnetic susceptibility  at different energy gaps $\epsilon$ and external fields $h$
 (numbers of curves correspond to the same values of  $h$ as in figure~\ref{fig2}).
}
\label{fig3}
\end{figure}

Dependencies of magnetic susceptibility $\chi$ (in arbitrary units) on the temperature $t$ for different values of magnetic fields $h$ and energy gap $\epsilon$ are presented in figure~\ref{fig3}. Naturally, for $h < h_\textrm{c}$, its behaviour is typical of the systems with the first order phase transition point. At $h = h_\textrm{c}$, the curves transform into the characteristic for a system with the second order phase transition point. For $h > h_\textrm{c}$, the magnetic susceptibility behaviour is characteristic of paramagnetic systems.

For $\epsilon > 0.5$ instance, only one curve $(h=0)$ demonstrates the behaviour typical of the systems with a possible phase transition in the point $T=0$~K [figure~\ref{fig3}~(d)]. Spontaneous magnetization here is impossible at any finite temperature, but the system is sensitive to an infinitesimally small external field at $T\rightarrow 0$~K.

\section{Thermodynamic functions}

Using the expression (\ref{eq3.5}) we can rewrite the equation (\ref{eq2.9}) for the free energy in the following form:
\be
\label{eq4.1}
F = - \frac{1}{\beta} \ln \frac{2}{(1-c)\sqrt{1-s^2}}  + \frac{3}{2} c^2 s^2 X_v J.
\ee
This representation for $F$ takes into account the dependencies of $c$ and $s$ on $t$, $h$, $\epsilon$ and will be useful for the calculation of different thermodynamic functions.

Having differentiated free energy (\ref{eq4.1}) on temperature, we obtained for entropy the following expression:
\be
\label{eq4.2}
S =  N k \left[ \ln \frac{2}{(1-c)\sqrt{1-s^2}} + \lp 3c s^2 \beta X_v J - \frac{1}{1-c} \rp X_1 +
\lp 3c^2 s \beta X_v J - \frac{s}{1-s^2} \rp X_2 \right],
\ee
where
\bea
&&
X_1  =  \frac{2 c s (c^2 s + h) + (c s^2-\varepsilon_0) Y_2}{Y_1Y_2-4c^2s^2}\,, \qquad %\non &&
\label{eq4.3}
X_2  = \frac{(c^2 s + h) Y_1 + 2c s(c s^2-\varepsilon_0) }{Y_1Y_2-4c^2s^2}\,,
\eea
$Y_1, Y_2$ are determined by expressions (\ref{eq3.12}).

\begin{figure}[!t]
\centerline{
\includegraphics[width=0.43\textwidth]{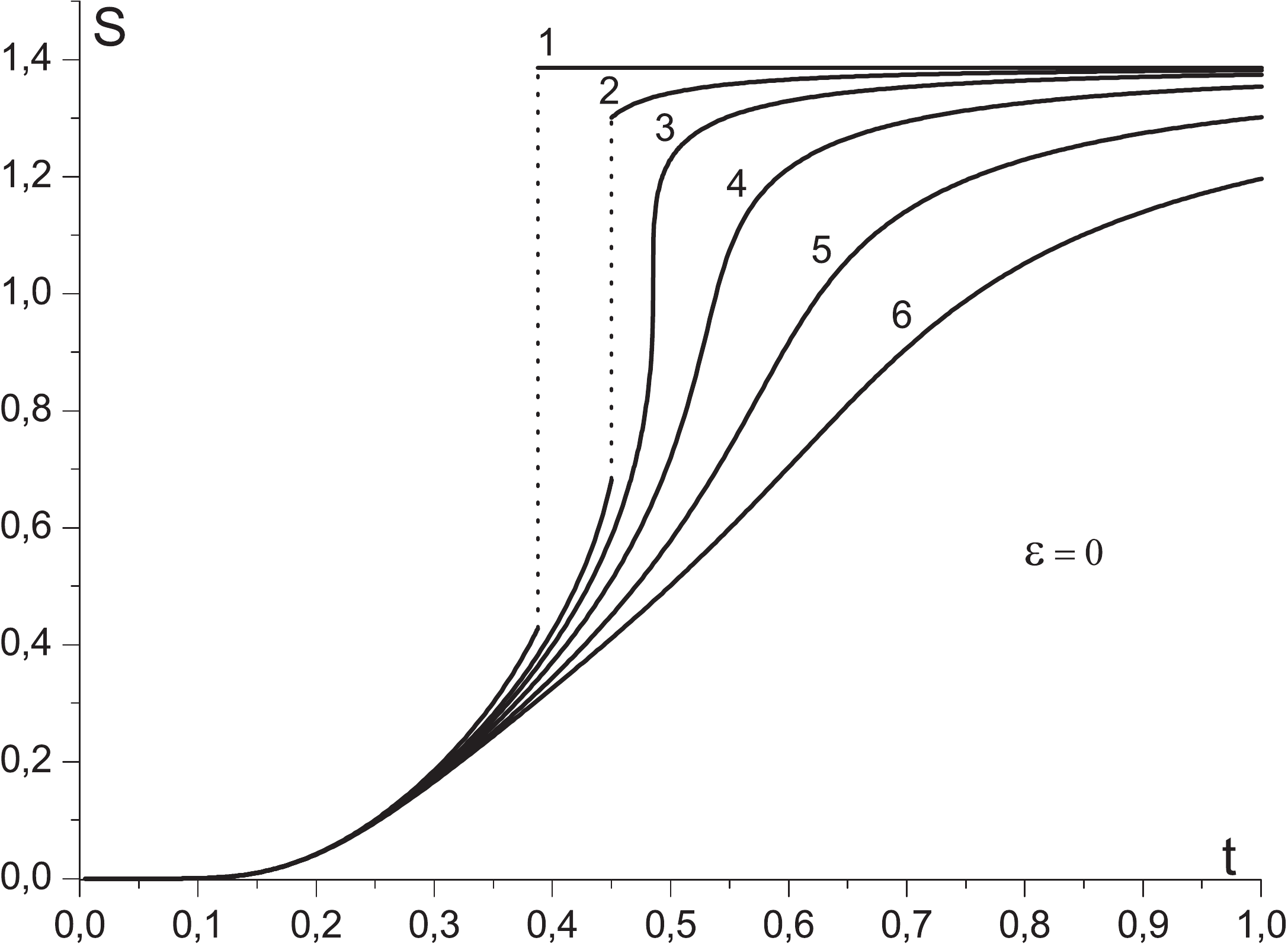}
\qquad
\includegraphics[width=0.43\textwidth]{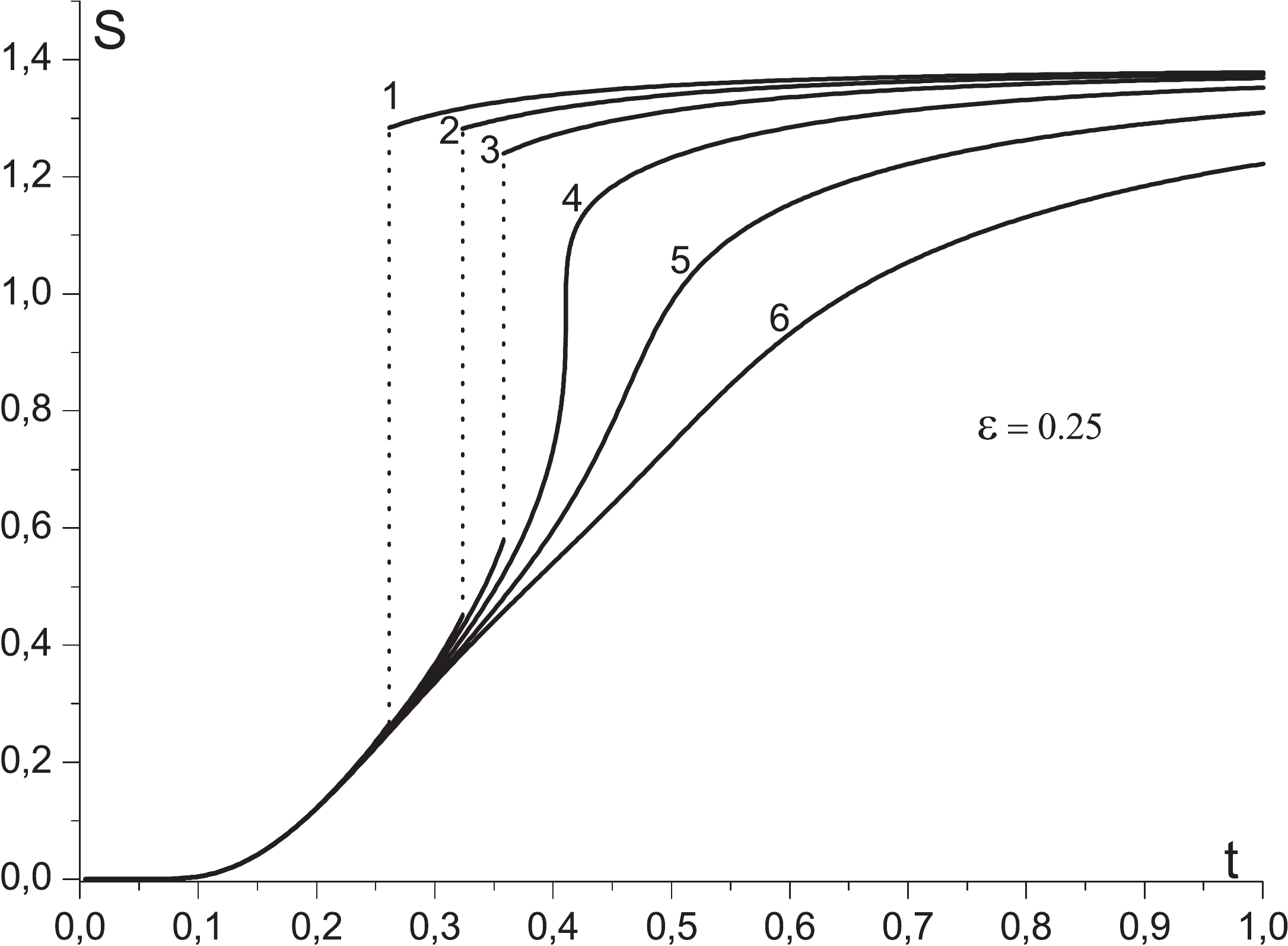}
}\centerline{(a) \hspace{0.45\textwidth} (b)}
\centerline{
\includegraphics[width=0.43\textwidth]{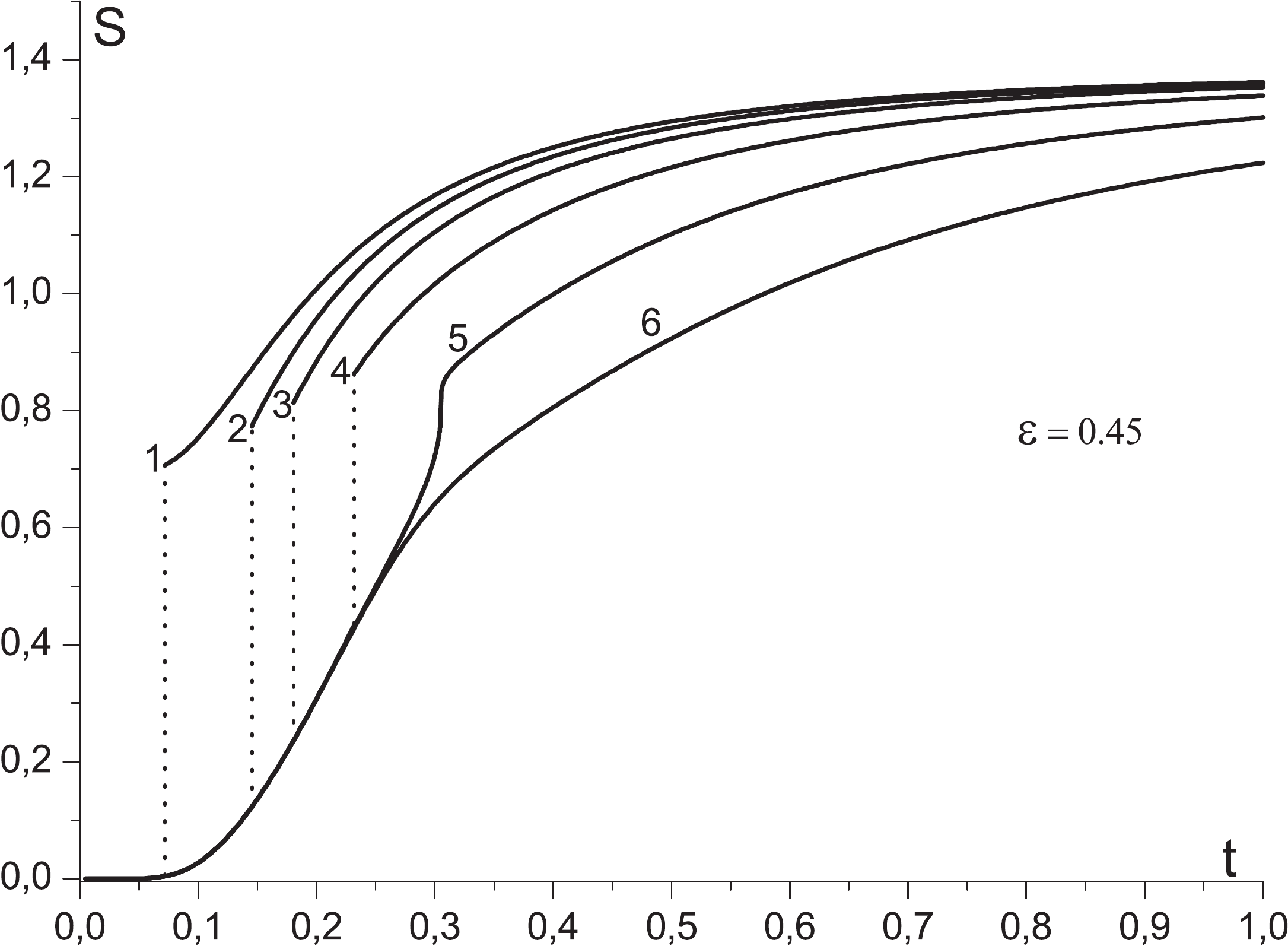}
\qquad
\includegraphics[width=0.43\textwidth]{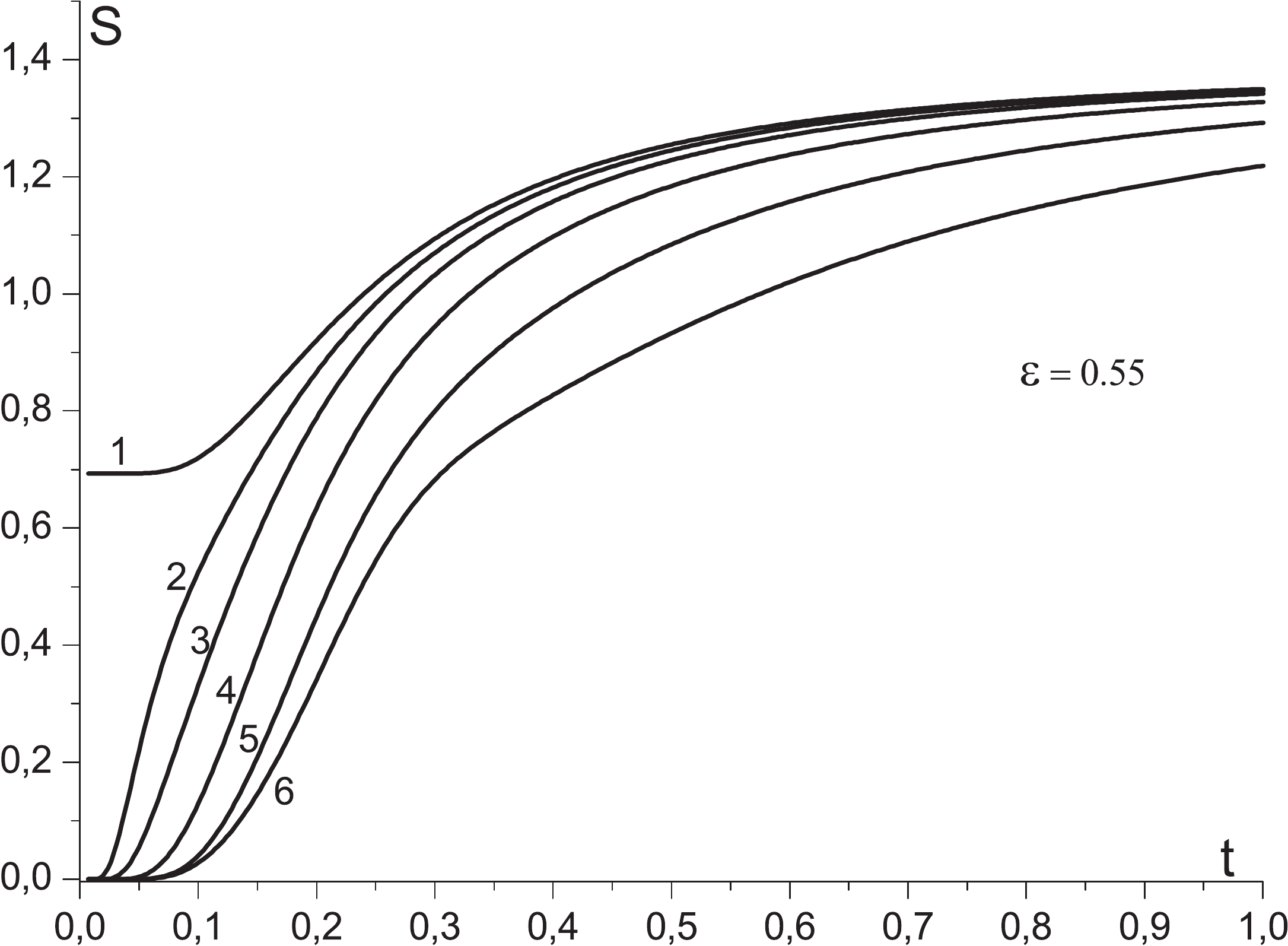}
}\centerline{(c) \hspace{0.45\textwidth} (d)}
\caption{
Temperature dependencies of the entropy at different energy gaps $\epsilon$ and external fields $h$ (numbers of curves correspond to the same values of  $h$ as in figure~\ref{fig2}).}
\label{fig4}
\end{figure}

The dependencies of entropy $S$ (in arbitrary units) on the reduced temperature at different values of external field $h$ and energy gap $\epsilon$ are presented in figure~\ref{fig4}. All of them demonstrate significant jumps at some temperature points dependent on the field for all $h<h_\textrm{c}$ [see figures~\ref{fig4}~(a)--(c)]. This confirms the existence of the first order magnetic phase transition. For $h=h_\textrm{c}$ there is a noticeable kink in the entropy behaviour in the corresponding temperature point. Such a behaviour is in accordance with the thermodynamic classification of the phase transition order \cite{ref29,ref30}.

A specific role of the zero external field at the $\epsilon>0.5$ instance is  demonstrated in figure~\ref{fig4}~(d). At $T\rightarrow 0$~K, the total entropy of the system in zero field ($h=0$) does not turn into zero, because magnetic ordering  is not realized. At the same time, all particles remain at the ground level ($c=0$) and the configuration part of entropy is equal to zero.

Comparison of numerical values of entropy to the right and to the left of the transition point makes it possible  to find the latent heat of the first order magnetic phase transition:
\be
\label{eq4.4}
q = T_\textrm{c} \lp S_r - S_l \rp,
\ee
where $S_r = S_+$ is a limit value of $S$ when temperature falls to the phase transition point $T_\textrm{c}$, and $S_l = S_-$ is a limit value of $S$ when temperature increases to this point.

\begin{figure}[!t]
\centerline{
\includegraphics[width=0.5\textwidth]{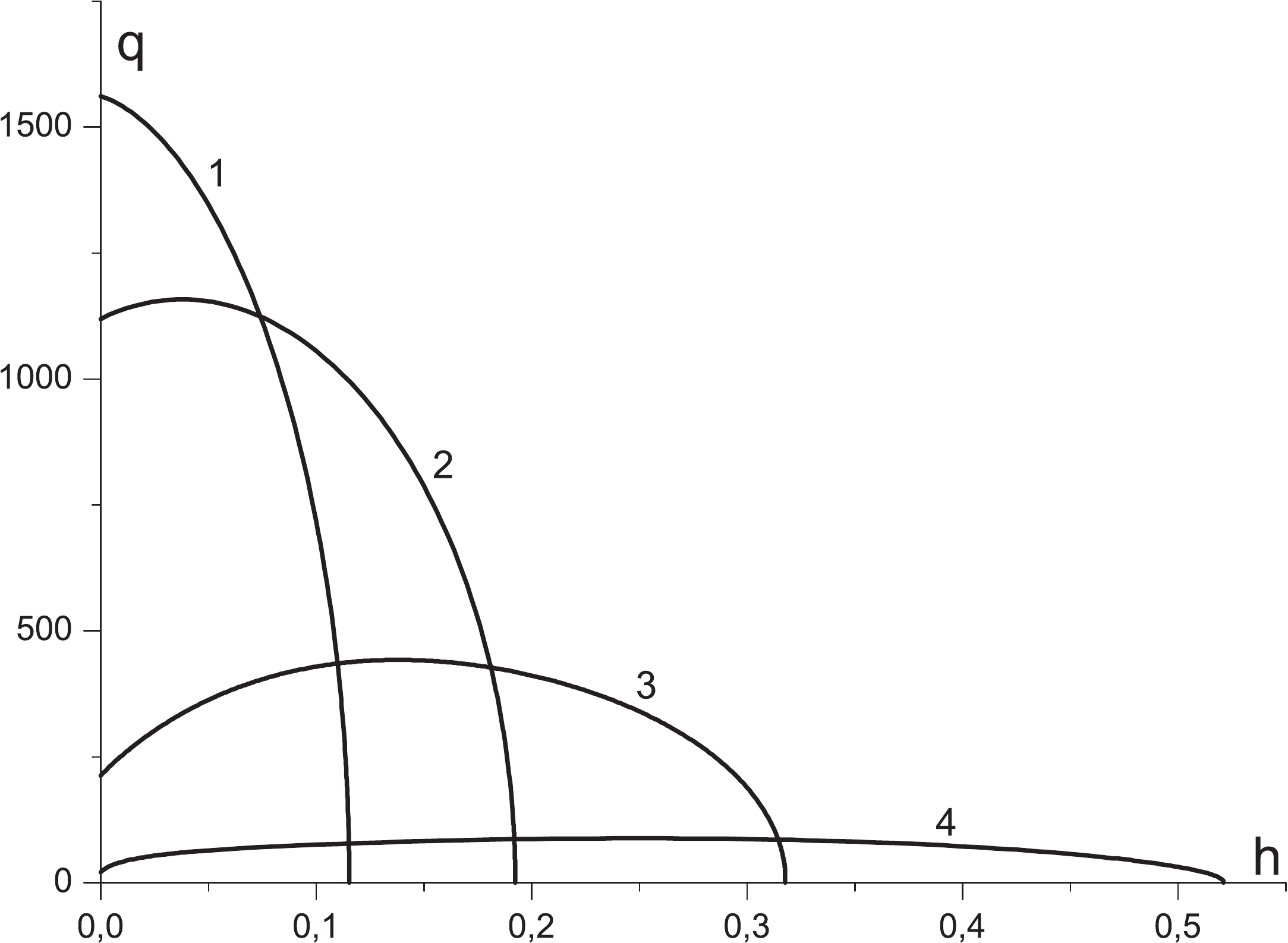}
}
\caption{
External field dependencies of latent heat at different energy gaps $c$  (curves: 1~--- $\epsilon=0$, 2~--- $\epsilon=0.25$, 3~--- $\epsilon=0.45$, 4~--- $\epsilon=0.495$).}
\label{fig5}
\end{figure}

The latent heat (in arbitrary units) behaviour under the external field $h$ is presented in figure~\ref{fig5}. The $q$ is not equal to zero for $h < h_\textrm{c}$ and vanishes at $h = h_\textrm{c}$ for all $\epsilon$. Every curve possesses a slight maximum connected, in our opinion, with the different temperature behaviours of $c$ and $s$ parameters.

\begin{figure}[!b]
\centerline{%
\includegraphics[width=0.43\textwidth]{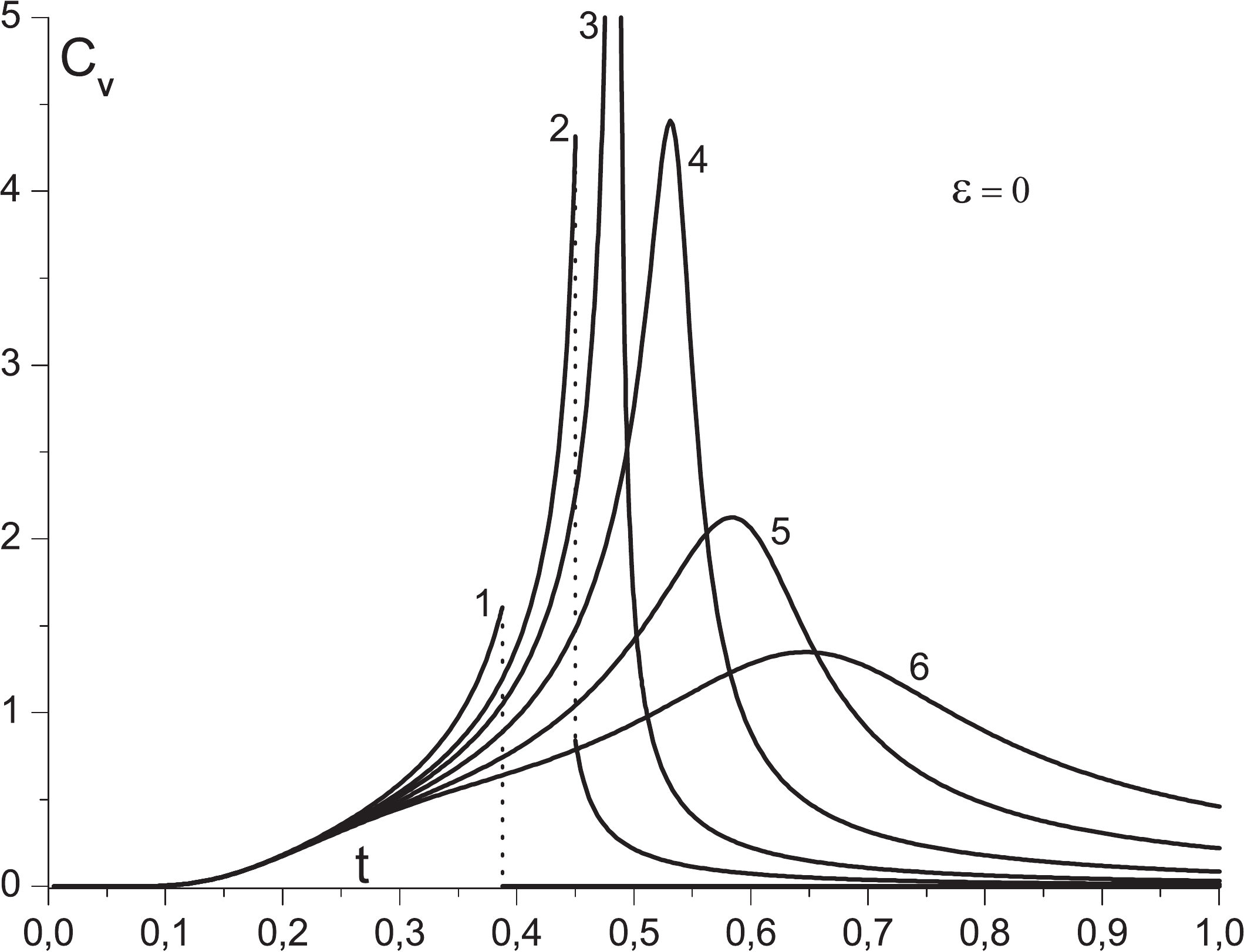}
\qquad
\includegraphics[width=0.43\textwidth]{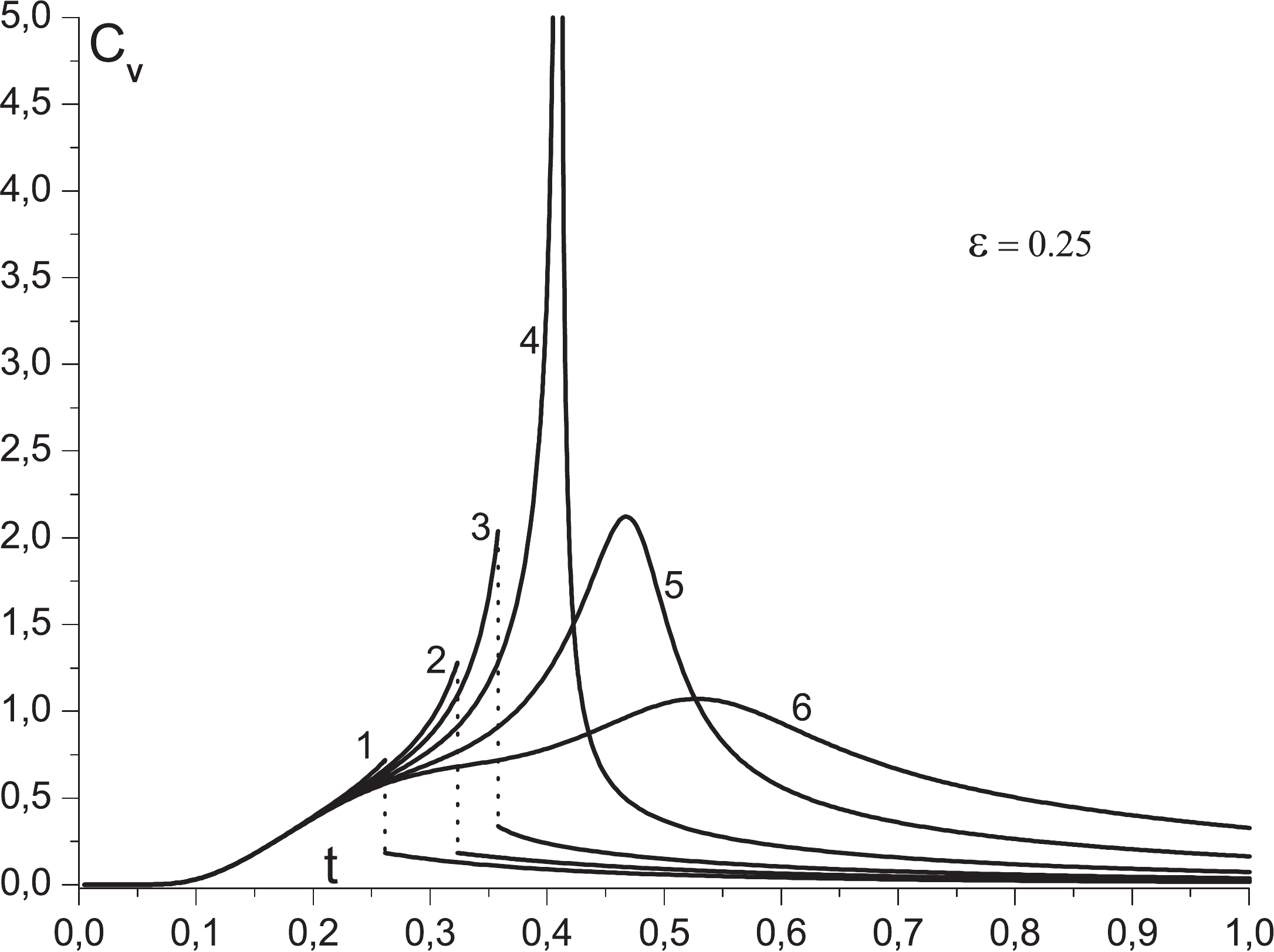}
}\centerline{(a) \hspace{0.45\textwidth} (b)}
\centerline{
\includegraphics[width=0.43\textwidth]{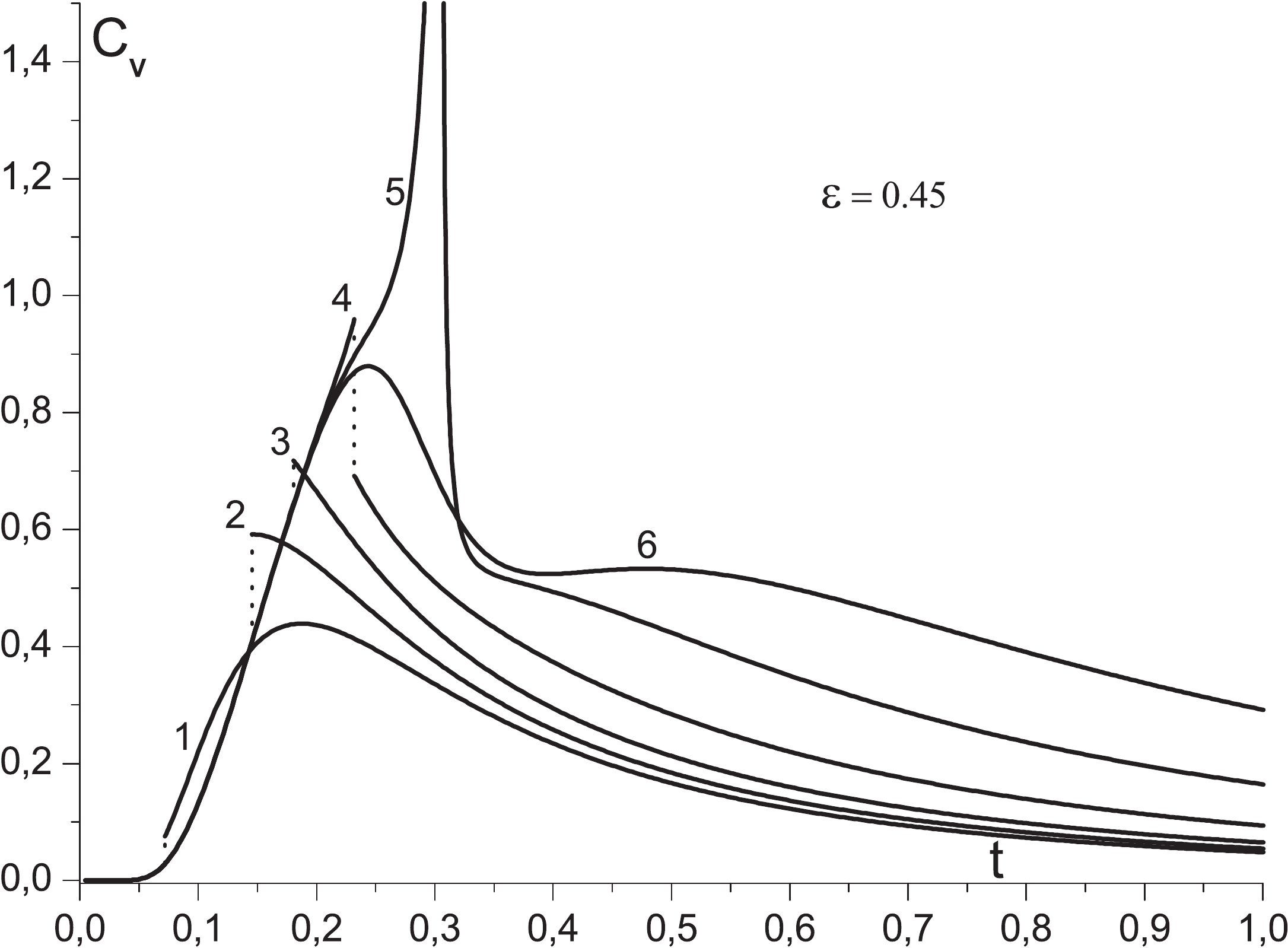}
\qquad
\includegraphics[width=0.43\textwidth]{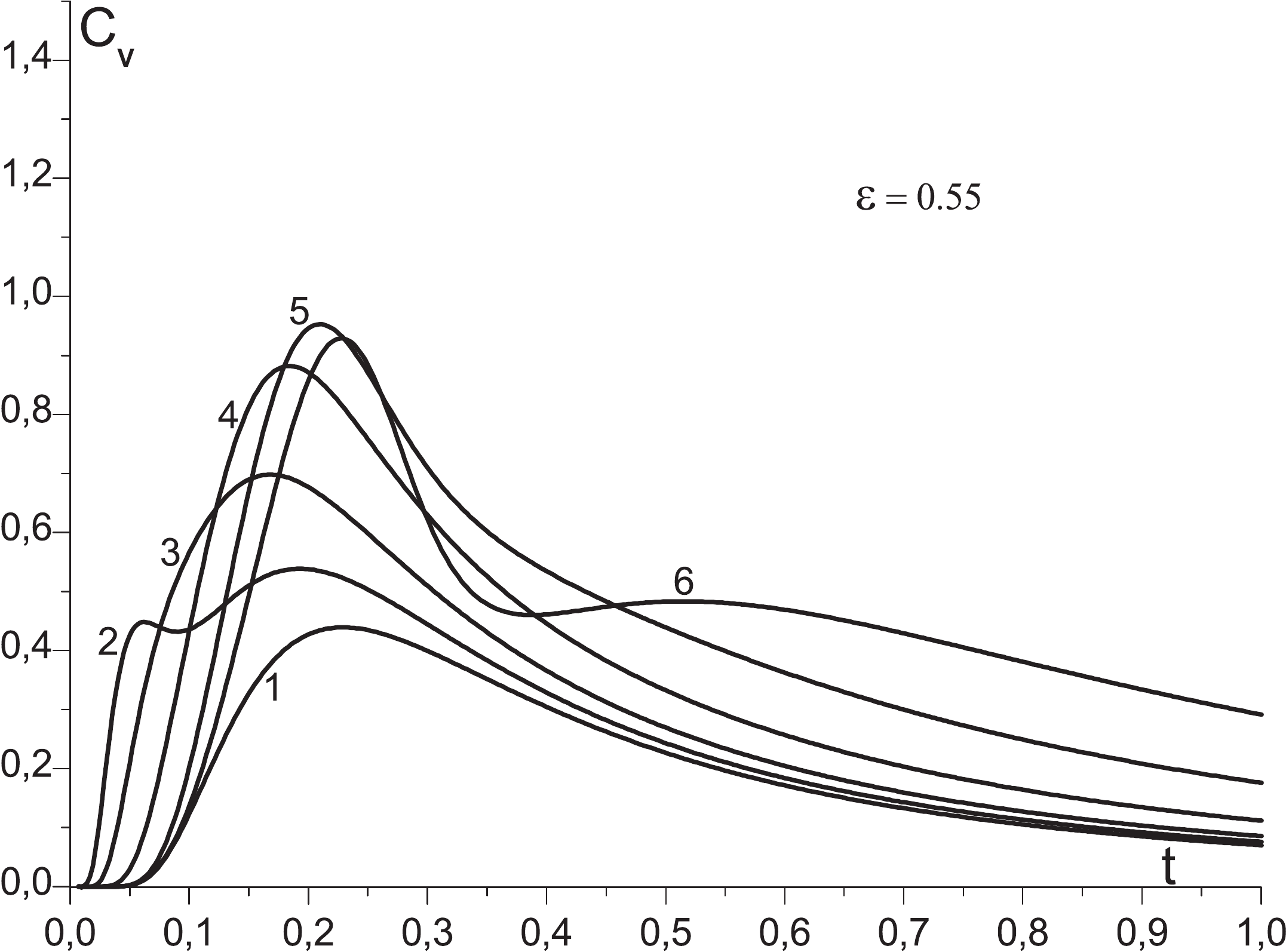}
}\centerline{(c) \hspace{0.45\textwidth} (d)}
\caption{
 Temperature dependencies of the heat capacity at different energy gaps $\epsilon$ and external fields $h$
 (numbers of curves correspond to the same values of  $h$ as in figure~\ref{fig2}).}
\label{fig6}
\end{figure}

Based on the well-known thermodynamic relation:
\be
\label{eq4.5}
C_V = - T \lp \frac{\partial^2 F}{\partial T^2}\rp
\ee
the heat capacity (specific heat) has been calculated.
Being cumbersome the expression for $C_V$ is omitted here. The temperature dependencies of heat capacity (in arbitrary units) at different values of  external field $h$ at some $\epsilon$ are presented in figure~\ref{fig6}. All of them demonstrate an infinite growth at temperatures of the first and the second order phase transitions and a smooth behaviour for $h>h_\textrm{c}$ ($\epsilon<0,5$) [see figures~\ref{fig6}~(a)--(c)].

A specific place of the $\epsilon>0.5$ instance [figure~\ref{fig6}~(d)] manifests itself in the appearance of an additional maximum of $C_V$  for small (curve~2) and large (curve~6) fields. As can be seen by comparison with figures~\ref{fig2}~(d) and \ref{fig2}~(h), the right-hand maximum in curve~2 and the left-hand maximum in curve~6 correspond to a rapid growth of occupation of the excited level by particles ($c$), but both supplementary maxima correspond to a rapid growth of magnetization ($s$). Thus, for $\epsilon>0.5$, the temperature behaviours of $c$ and $s$ are noticeably independent. Such a tendency is also visible for $\epsilon=0.45$ [see figures~\ref{fig6}~(c)]. It should be underlined that the capacity calculated here corresponds only to the site excited and spin orientation degrees of freedom. Other important degrees of freedom, vibrational or rotational, for example, are not taken into account here.

The points of a visible change of thermodynamic functions behaviour (jumps for order parameters $c$ and $s$) are connected with a transition of the free energy (\ref{eq4.1}) from a certain branch to another one. The numerical analysis shows the existence of 1--3 different branches of $F$ in the whole temperature range from 0 to $\infty$. Two of them realized a minimum value of $F$ and the third one realized its maximum value. Naturally, the real behaviour of the system investigated is determined by the branches with minimum $F$. The temperature points of critical behaviour are found from the continuity condition of free energy (\ref{eq4.1}) in those points [$F(T_\textrm{c}, c_1, s_1)=F(T_\textrm{c}, c_2, s_2)$]:
\be
\label{eq4.6}
\ln \frac{(1-c_1)\sqrt{1-s_1^2}}{(1-c_2)\sqrt{1-s_2^2}} = \frac{3}{2} \beta_\textrm{c} X_v J \left( c_2^2 s_2^2 - c_1^2 s_1^2\right),
\ee
where $c_1$, $s_1$ and $c_2$, $s_2$ are roots of equations (\ref{eq3.5}) at $\beta\rightarrow\beta_\textrm{c}$ for two different branches of $F$ and $\beta_\textrm{c} = (kT_\textrm{c})^{-1}$.

The analysis of all physical properties of the system investigated  shows their strict connection with the behaviour of $c$ and $s$ parameters. Rapid changes of thermodynamic functions take place when the occupation number $c$ jumps from a value above 0.5 to a value below 0.5. According to equations (\ref{eq3.5}), this corresponds to a relation:
\be
\label{eq4.7}
c s^2 - \epsilon = 0.
\ee
Consequently, the excitation energy gap $\epsilon=0.5$ determines the upper limit of $\epsilon$, above which all changes in a system have a continuous character.

It is hard to compare the theoretical results obtained here with the other ones due to the lack of an exact coincidence of our model with the one well-known in the literature. However, experimental investigations of magnetic phenomena induced by thermal or photo excited electrons in complex metal or metal-organic species are now intensively carried out \cite{ref31,ref32,ref33,ref34}. Photo-induced magnetic structures based on the organic matrix intercalated by Rb and Cs atoms demonstrate thermodynamic properties above the Curie temperature close to the ones presented in figure~\ref{fig3} and figure~\ref{fig6}. Those structures in general are stable at low temperatures and only in V-Cr Prussian Blue Analogy compounds instance \cite{ref34}, a critical temperature for magnetization state is about 350~K. Molecular based magnets of this type are a relatively new class of materials which have good prospects to be used as spin-electronics memory cells. The physical mechanisms of their operation can be quite fully explained within the framework of the two-level four state model discussed here.

\section{Conclusions}

A self-consistent theory for a description of the excited energy levels occupation and spin-orientation ordering in the crystalline lattice is proposed. The Ising-like magnetic interaction between particles in the excited states are taken into account. When this interaction is omitted, the occupation number $c$ and the spin-order parameter $s$ become continuous functions of temperature and an external field, which is characteristic of two independent subsystems: noninteracting two-level site system and diluted magnetic system.

The exchange interaction being taken into account significantly changes the situation. Depending on the excited energy gap and on the magnitude of the external magnetic field, the transition to a spontaneous magnetically ordered state of the system is possible only for the fields less than the critical one. Moreover, at the field less than critical, there  holds the first order phase transition, and when this field turns into a critical value, the transition becomes the second order. For the fields bigger than the critical one, the spontaneous magnetization  of a system is suppressed by an external field (an induced magnetization is proportional to the field magnitude).

Those effects take place for the reduced excited energy gap $\epsilon\equiv\epsilon_0(X_v J)^{-1}$ less than 0.5. For $\epsilon\geqslant 0.5$, no spontaneous magnetic ordering is possible, total magnetization is completely determined by the  external magnetic field.

The generalized  (dependent on the external field) magnetic susceptibility $\chi$ behaviour is in accordance with the microscopic parameters $c$ and $s$ temperature dependencies. Its infinite peak at the phase transition points, for fields less than and equal to the critical fields, moves to a significant peak at higher fields. The width of $\chi$ curve near $T_\textrm{c}$ is much smaller for the first order phase transition instance ($h < h_\textrm{c}$) compared to the second order phase transition instance $(h = h_\textrm{c})$.

The latent heat $q$ of the first order phase transition in the system investigated is strongly field dependent, vanishing for $h \geqslant h_\textrm{c}$ in accordance with the change of the order of the phase transition. The heat capacity (specific heat) behaves in a usual manner and for $h \geqslant h_\textrm{c}$ remains finite for all temperatures.

All those peculiarities for $\chi$, $q$, $C_V$ depend on the $\epsilon$ value in accordance with  the above mentioned  objections. A specific place of the excited level gap $\epsilon=0.5$ is established as a limit value between the jump-like and continuous behaviour of thermodynamic functions of the investigated system.

Based on all of the research, we may conclude that in the investigated system, correlation between the excited level occupation and magnetic ordering plays a decisive role. Specific percolation effects concerning the magnetic properties of a system take place. They manifest themselves in a self-consistent way.  Disappearance of magnetic ordering entails a sharp decrease in the probability of occupation of the excited levels by particles.

The additional stimuli (optical radiation for example) being taken into account for particle excitation are quite promising in the study of the  mechanisms of photo-induced magnetic phenomena, which are candidates for future applications in spin-electronics.

The  proposed model also possesses other peculiarities based on the complex ratio between the excited level gap ($\epsilon$) and the magnitude of the external field ($h$). A detailed study of these peculiarities will be a matter of future investigations.

\section*{Acknowledgements}

Authors thank Prof. I.V.~Stasyuk for his interest to the present paper and for useful discussions of the subject of this research.

%\newpage

%%%%%%%%%%%%%%%%%%%%%%%%%%%%%%%%%%%%%%%%%%%%%%%%%%%%%%%%%%%%%%%%%%%%%%%%%%%%%%%%%%%%%%%%%%%

\ukrainianpart

\title{
Самоузгоджений підхід до процесів магнітного впорядкування та заповнення збуджених станів \\ у дворівневій системі}

\author[М.А. Кориневський, В.Б. Солов'ян]{М.А. Кориневський\refaddr{label1,label2,label3},
        В.Б. Солов'ян\refaddr{label1}}
\addresses{
\addr{label1} Інститут фізики конденсованих систем НАН України, вул. І.~Свєнціцького, 1, 79011 Львів, Україна
\addr{label2} Національний університет ``Львівська політехніка'', вул. С.~Бандери, 12, 79013 Львів, Україна
\addr{label3} Інститут фізики Щецінського університету, вул. Вєлькопольська, 15, 70451 Щецін, Польща
}

\makeukrtitle

\begin{abstract}
\tolerance=3000%
Детально вивчено процес феромагнітного впорядкування у дворівневій частково збудженій системі.
Величина намагніченості (магнітний параметр
впорядкування) та граткове впорядкування (заповнення збуджених рівнів) розраховані самоузгоджено.
Обговорено вплив зовнішнього магнітного поля і ширини енергетичної щілини між основним і збудженим рівнями на феромагнітний фазовий перехід.
\keywords магнетики, дворівнева система, фазовий перехід
\end{abstract}

\end{document}